\documentclass[aip,amsmath,amssymb,reprint,]{revtex4-1}

\usepackage{graphicx}% Include figure files
\usepackage{dcolumn}% Align table columns on decimal point
\usepackage{bm}% bold math
\usepackage[utf8]{inputenc}
\usepackage[T1]{fontenc}
\usepackage{mathptmx}
\usepackage{etoolbox}
\usepackage{xcolor}
\usepackage{hyperref}

\usepackage{verbatim}
\makeatletter
\def\@email#1#2{%
 \endgroup
 \patchcmd{\titleblock@produce}
  {\frontmatter@RRAPformat}
  {\frontmatter@RRAPformat{\produce@RRAP{*#1\href{mailto:#2}{#2}}}\frontmatter@RRAPformat}
  {}{}
}%
\makeatother

\begin{document}

% Don't count these!
%TC:ignore
%\quickwordcount{aipsamp}
%\quickcharcount{aipsamp}
%\detailtexcount{aipsamp} %<- this one
%TC:endignore

\preprint{AIP/123-QED}

%TC:ignore

\title[Spontaneous generation of athermal phonon bursts within bulk silicon causing excess noise, low energy background events and quasiparticle poisoning in superconducting sensors]{Spontaneous generation of athermal phonon bursts within bulk silicon causing excess noise, low energy background events and quasiparticle poisoning in superconducting sensors}

% Force line breaks with \\

%\author{R. Anthony-Petersen} \affiliation{University of California Berkeley, Department of Physics, Berkeley, CA 94720, USA}
%\author{A. Biekert} \affiliation{University of California Berkeley, Department of Physics, Berkeley, CA 94720, USA}
%\author{H. Birch} \affiliation{University of Michigan, Randall Laboratory of Physics, Ann Arbor, MI 48109-1040, USA}
%\author{T.K. Bui} \affiliation{International Center for Quantum-field Measurement Systems for Studies of the Universe and Particles (QUP,WPI), High Energy Accelerator Research Organization (KEK), Oho 1-1, Tsukuba, Ibaraki 305-0801, Japan}
\author{C.L. Chang} \affiliation{Argonne National Laboratory, 9700 S Cass Ave, Lemont, IL 60439, USA} \affiliation{Kavli Institute for Cosmological Physics, The University of Chicago, Chicago, IL 60637} \affiliation{Department of Astronomy and Astrophysics, The University of Chicago, Chicago, IL 60637}
\author{Y.-Y. Chang} \affiliation{University of California Berkeley, Department of Physics, Berkeley, CA 94720, USA}
%\author{L. Chaplinsky} \affiliation{University of Massachusetts, Amherst Center for Fundamental Interactions and Department of Physics, Amherst, MA 01003-9337 USA}
%\author{G. Cline} \affiliation{Lawrence Berkeley National Laboratory, 1 Cyclotron Rd., Berkeley, CA 94720, USA}
%\author{A. Dushkin} \affiliation{University of Michigan, Randall Laboratory of Physics, Ann Arbor, MI 48109-1040, USA}
%\author{C.W. Fink} \affiliation{University of California Berkeley, Department of Physics, Berkeley, CA 94720, USA} \affiliation{Now at Los Alamos National Laboratory, Los Alamos, NM 87545}
%\author{S. Fiorucci} \affiliation{Lawrence Berkeley National Laboratory, 1 Cyclotron Rd., Berkeley, CA 94720, USA}
\author{M. Garcia-Sciveres} \affiliation{Lawrence Berkeley National Laboratory, 1 Cyclotron Rd., Berkeley, CA 94720, USA} \affiliation{International Center for Quantum-field Measurement Systems for Studies of the Universe and Particles (QUP,WPI), High Energy Accelerator Research Organization (KEK), Oho 1-1, Tsukuba, Ibaraki 305-0801, Japan}
%\author{G. Gilchriese} \affiliation{Lawrence Berkeley National Laboratory, 1 Cyclotron Rd., Berkeley, CA 94720, USA}
\author{W. Guo} \affiliation{Department of Mechanical Engineering, FAMU-FSU College of Engineering, Florida State University, Tallahassee, FL 32310, USA} \affiliation{National High Magnetic Field Laboratory, Tallahassee, FL 32310, USA}
\author{S.A. Hertel} \affiliation{University of Massachusetts, Amherst Center for Fundamental Interactions and Department of Physics, Amherst, MA 01003-9337 USA}
%\author{A. Jastram} \affiliation{Texas A\&M University, Department of Physics and Astronomy, College Station, TX 77843-4242, USA}
\author{X. Li} \affiliation{Lawrence Berkeley National Laboratory, 1 Cyclotron Rd., Berkeley, CA 94720, USA}
\author{J. Lin} \affiliation{University of California Berkeley, Department of Physics, Berkeley, CA 94720, USA} \affiliation{Lawrence Berkeley National Laboratory, 1 Cyclotron Rd., Berkeley, CA 94720, USA}
\author{M. Lisovenko} \affiliation{Argonne National Laboratory, 9700 S Cass Ave, Lemont, IL 60439, USA}
\author{R. Mahapatra} \affiliation{Texas A\&M University, Department of Physics and Astronomy, College Station, TX 77843-4242, USA}
\author{W. Matava} \affiliation{University of California Berkeley, Department of Physics, Berkeley, CA 94720, USA}
\author{D.N. McKinsey} \affiliation{University of California Berkeley, Department of Physics, Berkeley, CA 94720, USA} \affiliation{Lawrence Berkeley National Laboratory, 1 Cyclotron Rd., Berkeley, CA 94720, USA}
%\author{D.Z. Osterman} \affiliation{University of Massachusetts, Amherst Center for Fundamental Interactions and Department of Physics, Amherst, MA 01003-9337 USA}
\author{P.K. Patel} \affiliation{University of Massachusetts, Amherst Center for Fundamental Interactions and Department of Physics, Amherst, MA 01003-9337 USA}
\author{B. Penning} \affiliation{University of Zurich, Department of Physics, 8057 Zurich, Switzerland}
\author{H.D. Pinckney} \affiliation{University of Massachusetts, Amherst Center for Fundamental Interactions and Department of Physics, Amherst, MA 01003-9337 USA}
\author{M. Platt} \affiliation{Texas A\&M University, Department of Physics and Astronomy, College Station, TX 77843-4242, USA}
\author{M. Pyle} \affiliation{University of California Berkeley, Department of Physics, Berkeley, CA 94720, USA}
\author{Y. Qi} \affiliation{Department of Mechanical Engineering, FAMU-FSU College of Engineering, Florida State University, Tallahassee, FL 32310, USA} \affiliation{National High Magnetic Field Laboratory, Tallahassee, FL 32310, USA}
\author{M. Reed} \affiliation{University of California Berkeley, Department of Physics, Berkeley, CA 94720, USA}
\author{I. Rydstrom} \affiliation{University of California Berkeley, Department of Physics, Berkeley, CA 94720, USA}
%\author{G.R.C Rischbieter} \affiliation{University of Michigan, Randall Laboratory of Physics, Ann Arbor, MI 48109-1040, USA}
\author{R.K. Romani} \thanks{Corresponding author: \href{mailto:rkromani@berkeley.edu}{rkromani@berkeley.edu}}\affiliation{University of California Berkeley, Department of Physics, Berkeley, CA 94720, USA}
\author{B. Sadoulet}\affiliation{University of California Berkeley, Department of Physics, Berkeley, CA 94720, USA}
\author{B. Serfass} \affiliation{University of California Berkeley, Department of Physics, Berkeley, CA 94720, USA}
%\author{R.J.  Smith} \affiliation{University of California Berkeley, Department of Physics, Berkeley, CA 94720, USA}
\author{P. Sorensen} \affiliation{Lawrence Berkeley National Laboratory, 1 Cyclotron Rd., Berkeley, CA 94720, USA}
\author{B. Suerfu} \affiliation{International Center for Quantum-field Measurement Systems for Studies of the Universe and Particles (QUP,WPI), High Energy Accelerator Research Organization (KEK), Oho 1-1, Tsukuba, Ibaraki 305-0801, Japan}
%\author{A. Suzuki} \affiliation{Lawrence Berkeley National Laboratory, 1 Cyclotron Rd., Berkeley, CA 94720, USA}
\author{V. Velan} \affiliation{Lawrence Berkeley National Laboratory, 1 Cyclotron Rd., Berkeley, CA 94720, USA}
\author{G. Wang} \affiliation{Argonne National Laboratory, 9700 S Cass Ave, Lemont, IL 60439, USA}
\author{Y. Wang} \affiliation{University of California Berkeley, Department of Physics, Berkeley, CA 94720, USA} \affiliation{Lawrence Berkeley National Laboratory, 1 Cyclotron Rd., Berkeley, CA 94720, USA}
%\author{S.L. Watkins} \affiliation{University of California Berkeley, Department of Physics, Berkeley, CA 94720, USA}
\author{M.R. Williams} \affiliation{Lawrence Berkeley National Laboratory, 1 Cyclotron Rd., Berkeley, CA 94720, USA}
\author{V.G. Yefremenko} \affiliation{Argonne National Laboratory, 9700 S Cass Ave, Lemont, IL 60439, USA}

\collaboration{TESSERACT Collaboration}%\noaffiliation

 \email{rkromani@berkeley.edu.}

\date{\today}% It is always \today, today,
             %  but any date may be explicitly specified

\begin{abstract}
Solid state phonon detectors used in the search for dark matter and coherent neutrino nucleus interactions (CE$\nu$NS) require excellent energy resolution (eV-scale or below) and low backgrounds. An unknown source of phonon bursts, the low energy excess (LEE), dominates other above-threshold backgrounds and generates excess shot noise from sub-threshold bursts.  In this paper, we measure these phonon bursts for 12 days after cooldown in two nearly identical 1 cm$^2$ silicon detectors that differ only in the thickness of their substrate (1 mm vs. 4 mm thick). We find that both the channel-correlated shot noise and near-threshold shared LEE relax with time since cooldown. Additionally, both the correlated  shot noise and LEE rates scale linearly with substrate thickness. When combined with previous measurements of other silicon phonon detectors with different substrate geometries and mechanical support strategies, these measurements strongly suggest that the dominant source of both above and below threshold LEE is the bulk substrate. By monitoring the relation between bias power and excess phonon shot noise, we estimate that the energy scale for sub-threshold noise events is $0.68 \pm 0.38$ meV. In our final dataset, we report a world-leading energy resolution of 258.5$\pm$0.4 meV in the 1 mm thick detector. Simple calculations suggest that these silicon substrate phonon bursts are likely a significant source of quasiparticle poisoning in superconducting qubits operated in well shielded and vibration free environments.
\end{abstract}

\maketitle

%\linenumbers

\section{Motivation}
%TC:endignore

Solid state phonon detectors with excellent energy resolution and low backgrounds are a key technology with applications in fundamental physics and beyond. For example, the search for light dark matter candidates \cite{kuflikElasticallyDecouplingDark2016,kuflikPhenomenologyELDERDark2017a, hochbergMechanismThermalRelic2014,hochbergModelThermalRelic2015, hallFreezeinProductionFIMP2010} and the detection of rare eV-scale Coherent Elastic Neutrino Nucleus Scattering (CE$\nu$NS) from nuclear reactors \cite{Ricochet, Nucleus} require detectors with both low noise and low background rates to probe new regions of parameter space. High-resolution phonon sensors can also be used as a veto \cite{TESVeto} in photon-coupled rare event searches (e.g. dark photon or axion searches \cite{BREAD, LAMPPOST}).

Unfortunately, today's low-threshold phonon detectors observe orders of magnitude more background events below several hundred eV than expected from background radioactivity or cosmic ray interactions \cite{adariEXCESSWorkshopDescriptions2022, LEEReview}. These events are often referred to as the ``low energy excess'' (LEE). Measurements \cite{CRESSTTwoChannel, TwoChannelPaper} show that the LEE can be categorized into two distinct subclasses based on the channel energy partition in detectors with multi-channel phonon sensor readout. ``Single'' LEE events are found to deposit nearly all of their energy in a single phonon sensor channel, suggesting that they originate within the metal films of the phonon sensors themselves. By contrast, ``shared'' LEE events deposit nearly equal amounts of energy in all phonon sensor channels, which suggests a substrate origin. Similarly, our collaboration has observed both channel-correlated and uncorrelated excess noise in our phonon sensors, consistent with shot noise from sub-threshold events from these two classes \cite{TwoChannelPaper}. This work probes the scaling of these backgrounds and noise terms with substrate thickness to further understand their origin.

\section{Experimental Setup}

Here, we measure background events and noise in two nominally identical 1 cm$^2$ silicon low-threshold athermal phonon detectors, where the silicon substrate thickness was varied from 1 mm (mass: 0.233 grams) to 4 mm (mass: 0.932 grams). To read out phonon signals, the detectors use tungsten Transition Edge Sensors (TES, $T_c \approx$ 50 mK) coupled to aluminum phonon collection fins in the common Quasiparticle-trap-enhanced Electrothermal-feedback Transition Edge Sensor (QET) \cite{irwinQuasiparticleTrapAssisted1995} architecture, employing an identical design to the detectors in Ref. \cite{TwoChannelPaper}. These individual phonon sensors are aggregated into two separately read out channels to distinguish events and noise which deposit energy in a single phonon sensor from those that deposit energy in multiple phonon sensors \cite{TwoChannelPaper, CRESSTTwoChannel, TwoChannelTalkLTD}. The phonon sensors on these detectors have improved phonon energy collection efficiency, saturation energy, and channel uniformity compared to those from an earlier fabrication run that suffered from accidental tungsten over-etching \cite{TwoChannelPaper}. As in Ref. \cite{AnthonyPetersen2024}, both detectors are suspended by wire bonds to suppress backgrounds associated with detector holding and housed together in an IR and EMI shielded set of housings attached to the mixing chamber stage of a dilution refrigerator. They were read out using single-stage DC SQUID array amplifiers. 

To minimize LEE rate differences due to uncontrolled systematic variables, differences in fabrication, transport, storage, and measurement were minimized to the extent possible. Both detector substrates were detector-grade double side polished intrinsic silicon (70 k$\Omega$ cm for 1 mm thick, 20 k$\Omega$ cm for 4 mm thick). The detectors were fabricated in the same facility at Texas A\&M using identical procedures in consecutive fabrication runs separated by less than one week. Since LEE is known to potentially vary with time and thermal cycling, the detectors were transported, diced, stored, and run together in the same experimental setup at UC Berkeley using nominally identical electronic channels. Unfortunately, the detectors displayed different susceptibility to environmental vibrations, which we believe is due to uncontrolled variability in the wirebond hanging process. 

\begin{figure}
\includegraphics[width=1\columnwidth]{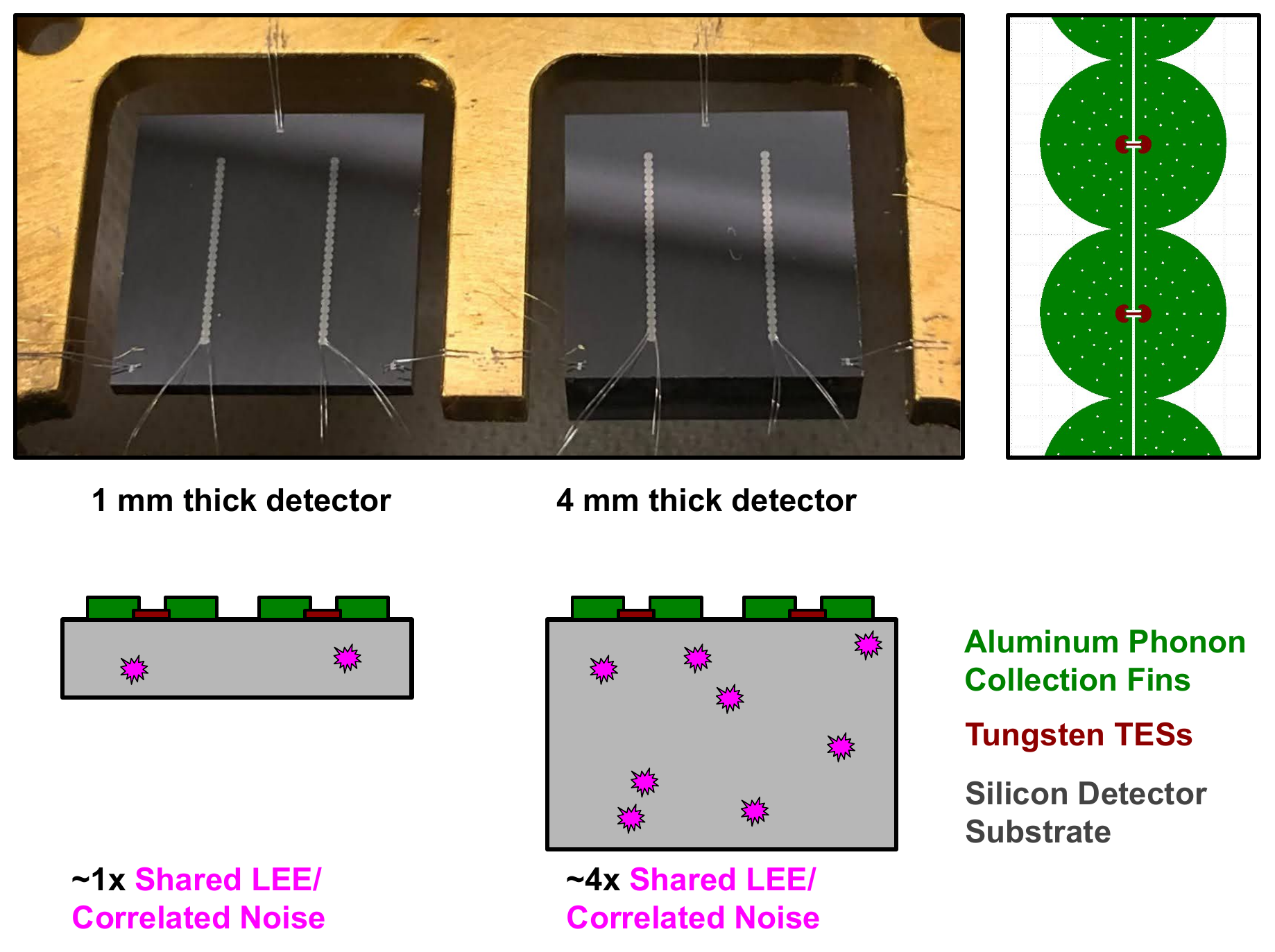}
\caption{\label{fig:diagram} (Top left) Photograph of our 1 mm (left) and 4 mm thick (right) 1 cm$^2$ silicon detectors. (Top right) Detail on mask design for our phonon sensors (QETs). For scale, the QET fins have a radius of approximately 140 $\mu$m. (Bottom) Sketch of backgrounds and shot noise sources we observe in our devices. Shared LEE backgrounds and correlated phonon shot noise appear at approximately 4x the rate in the 4 mm detector compared to the 1 mm detector.}
\end{figure}

\begin{figure}
\includegraphics[width=1\columnwidth]{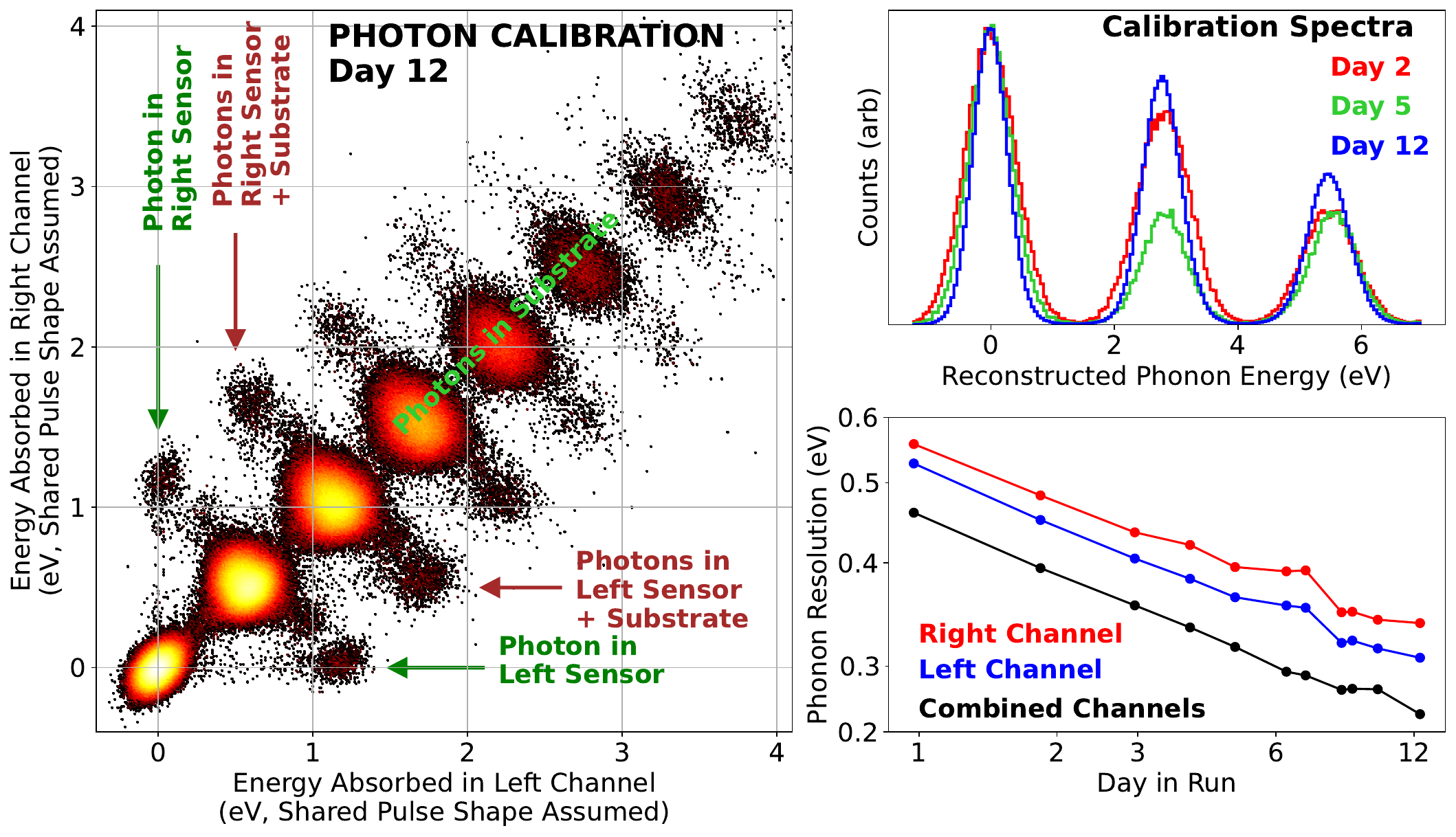}
\caption{\label{fig:calibration} 1 mm device calibration. (Left) Two dimensional histogram of calibration events. Most calibration photons hit the substrate, causing even responses in both channels (main diagonal events). Due to the energy resolution of our detector, individual photons appear as quantized quasi-Gaussian groups. Occasionally, photons hit a phonon sensor (QET), causing a large response in that channel (offset events). Around 1.1 eV of each 2.755 eV photon is absorbed in our sensors due to our phonon collection efficiency. (Top Right) Photon calibration spectra taken on days 2, 5, and 12. Note the improvement on baseline resolution over time. (Bottom Right) Measured phonon energy resolutions in the right, left, and combined channels over time. }
\end{figure}

To calibrate our detectors' response, we illuminate the phonon sensor face with pulses of 450 nm (2.755 eV) photons, emitted from a diffuser at the tip of a single mode fiber which runs from the detector cavity to a room temperature laser. To precisely record the time of photon calibration events, a logic level signal from the signal generator that powers the laser is concurrently recorded and stored with the continuously recorded signal data streams. Phonon signal amplitudes are then estimated offline using optimum filters\cite{TwoChannelPaper, TwoChannelLimits}. Plotting the height of the resulting events in each channel (Fig. \ref{fig:calibration} left), we observe the same features as in Ref. \cite{TwoChannelPaper}. Most photons are absorbed in the detector substrate system and produce approximately equal responses in both channels, while occasionally photons are directly absorbed by the aluminum phonon collection fins, producing a large saturated response in only one channel. Because multiple photons may strike the detector in one laser pulse, we also observe a superposition of these event types.

To combine the responses of the two channels of one detector, we use a ``2x1'' multichannel optimum filter, which simultaneously fits the response in both channels, scaling them together by one amplitude \cite{TwoChannelPaper, TwoChannelLimits}. In our final dataset, we achieve a baseline phonon energy resolution of 258.5$\pm$0.4 meV for this combined channel response, improving upon the resolution of our previously world-leading detector \cite{TwoChannelPaper, TwoChannelLimits} and prior work by other groups \cite{HVeVR4, CRESSTSoS}. 

We periodically calibrate our detector and find that the energy resolution of both detectors monotonically improves over the course of the run (Fig. \ref{fig:calibration} Bottom Right). In principle, this could be due to two factors: an improvement in the sensors' phonon collection efficiency over time, or a reduction in noise in the detector. Monitoring our detectors' phonon collection efficiencies over time, we find that they are approximately constant (see supplementary material section \ref{appendix:pce}), implying that our noise environment is improving with time.

\section{Measurement and Modeling of Detector Noise}

To characterize our detectors' noise over time, we record 2 hours of continuous data following each calibration dataset and remove time periods with abnormally low bias power, which is primarily caused by unusually high environmental vibrations or large energy depositions. Additionally, to select periods without above-threshold events that would contaminate and bias our noise, we find the largest magnitude pulse in the entire 200 ms period using an optimal filter and select periods for which this fit has a negative amplitude, i.e., the largest event in the trace is a statistical fluctuation rather than a real energy deposition. In total, 3.9\% of the 200 ms traces pass these very strict noise selection criteria and are used to estimate the noise Cross Power Spectral Density (CSD, see supplementary material section \ref{appendix:cuts} for further discussion of our cuts). We reference this CSD to noise equivalent power by applying a three-pole responsivity model\cite{ThreePoledPdI, irwinTransitionEdgeSensors2005} for each channel (i.e. $\partial P/ \partial I(f)$) that is estimated from IV and $\frac{\partial I}{\partial V}$ measurements as in Ref. \cite{TwoChannelPaper}.

\begin{figure}
\includegraphics[width=1\columnwidth]{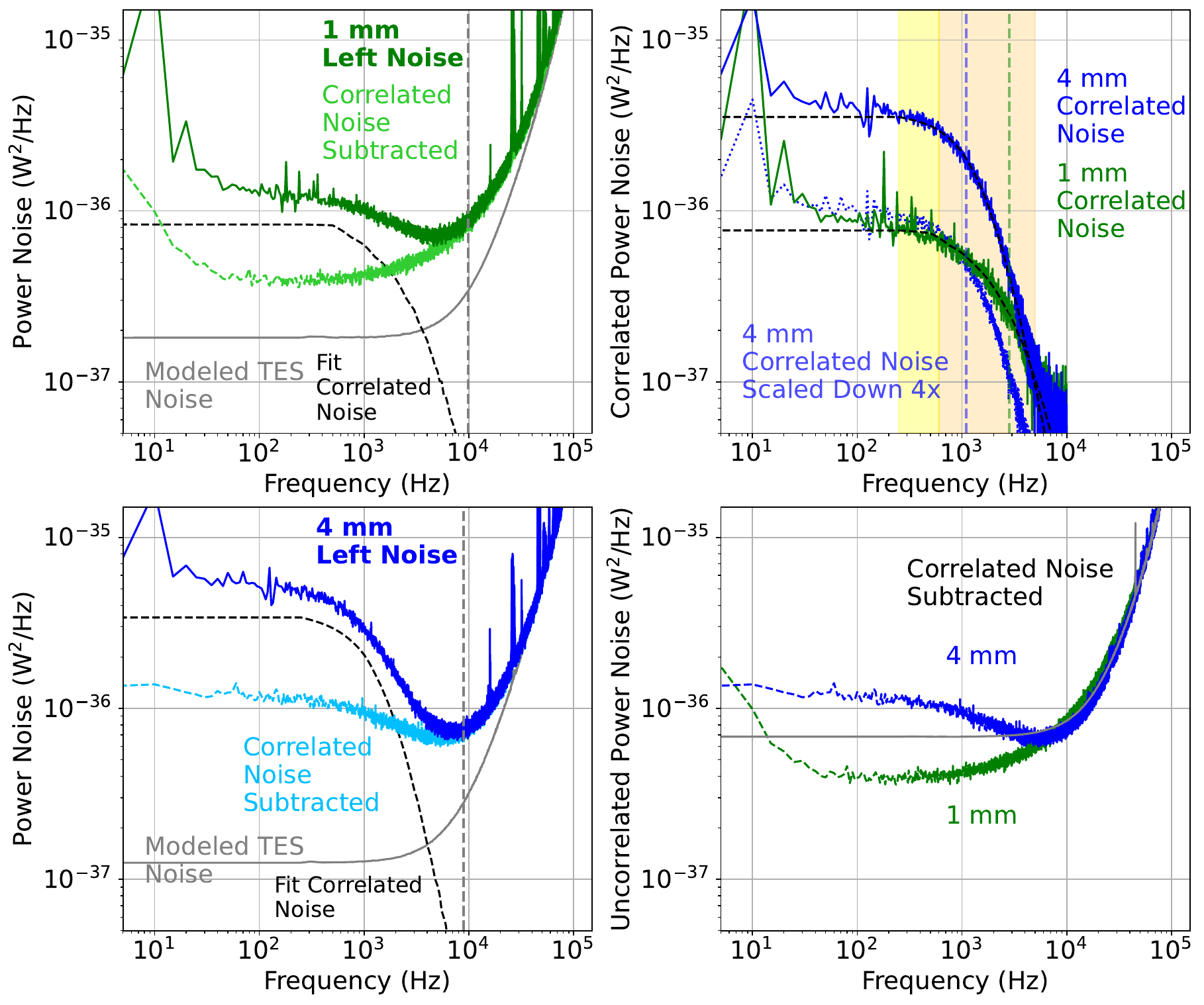}
\caption{\label{fig:noise} (Left) Noise in the left channels of the 1 mm (top, green) and 4 mm (bottom, blue) detectors. Our noise is well in excess of our modeled TES noise (gray, dotted), and is composed of correlated phonon noise and uncorrelated noise as in Ref. \cite{TwoChannelPaper}. The peaks around 150 Hz are due to vibration coupled noise. Grey dashed lines show the primary (electrothermal) pole of the TES. (Top Right) Correlated noise (i.e. off diagonal CSD element $S_{LR}$) in the 4 mm and 1 mm detectors. The correlated noise in the 4 mm detector is 4 times as large as the 1 mm detector. The amplitude of the 1 mm correlated noise is fit in the orange highlighted region (to avoid vibration coupled peaks), while the 4 mm noise is fit in the orange and yellow regions. Dashed green and blue lines show the primary phonon poles of the 1 mm and 4 mm detectors respectively. (Bottom Right) Uncorrelated noise (i.e. $S_{LL}$, with the modeled correlated noise subtracted) in the left channels of the 1 mm and 4 mm detectors, consistent with the modeled TES Johnson noise at high frequencies and an excess noise term that is approximately flat and consistent between the two channels at low frequencies (see section \ref{appendix:uncorrelated_noise} for further discussion). Data corresponds to the final (day 12) dataset.}
\end{figure}

Similar to our observations in Ref. \cite{TwoChannelPaper}, we find that the noise is significantly larger than the expected theoretical TES noise \cite{irwinTransitionEdgeSensors2005} and has a significant broad band channel correlated component (which should be statistically consistent with zero). The broad spectrum correlated noise is consistent with shot noise from many subthreshold phonon events in the substrate; i.e. it has the same frequency shape and channel energy partition ($\sim$50\% left, $\sim$50\% right) as substrate-interacting calibration photons (Fig. \ref{fig:noise} Top Right). The magnitude of this broad correlated noise appears to scale with the detector volume, since the correlated noise is four times larger in the 4 mm detector than in the 1 mm detector (Fig. \ref{fig:noise} Top Right).

After removing these correlated noise terms, we find that there is an additional quasi-flat uncorrelated noise term, in significant excess of the expected level of thermal fluctuation noise in the TESs, but in some tension with the flat shot noise model of Ref. \cite{TwoChannelPaper} (see supplementary material section \ref{appendix:uncorrelated_noise} for further discussion).

\begin{figure}
\includegraphics[width=1\columnwidth]{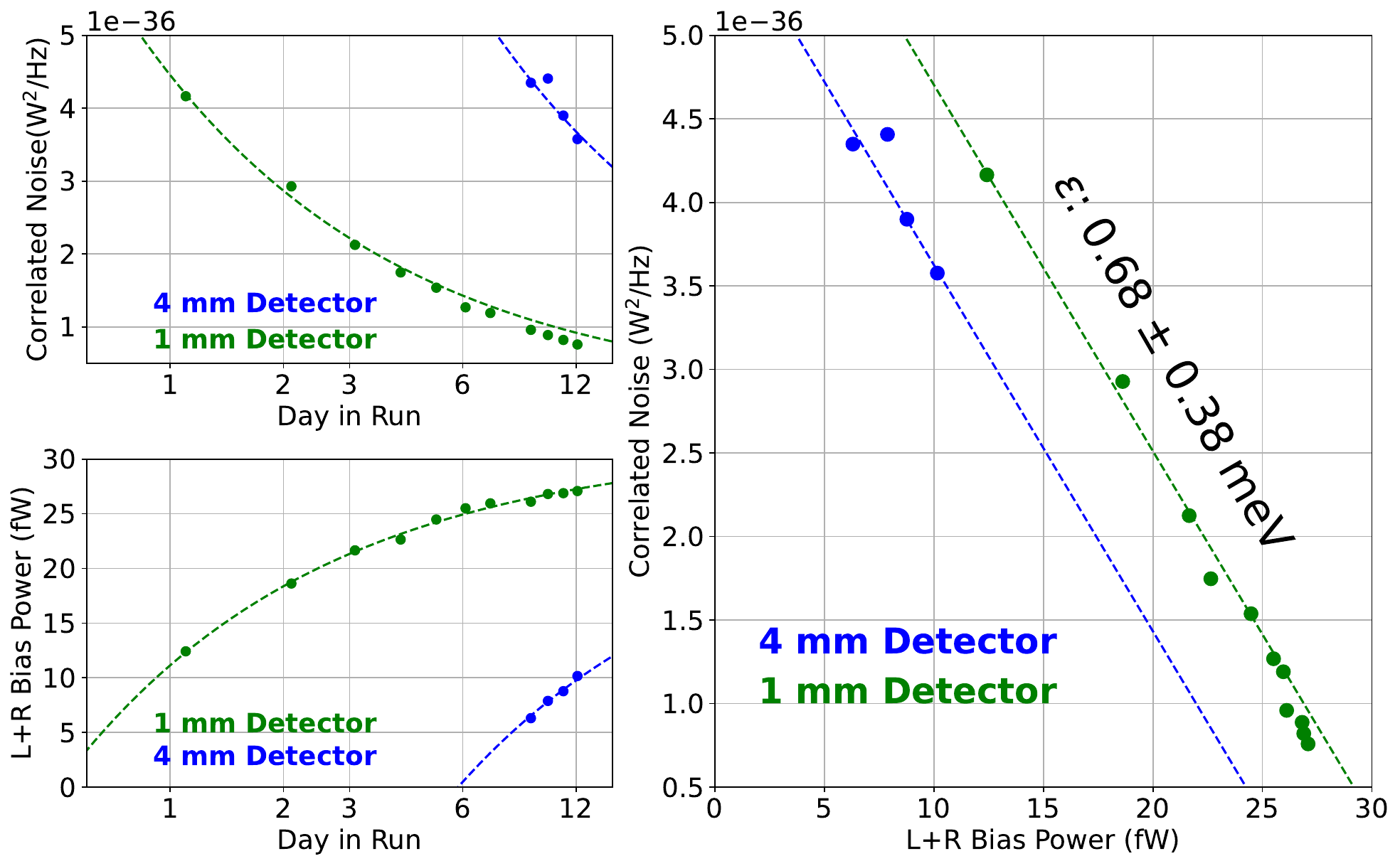}
\caption{\label{fig:noisevtime} Correlated noise level (Top Left) and Left + Right channel bias powers (Bottom Left) in the 1 mm (green) and 4 mm (blue) detectors as a function of time. (Right) Correlated noise level vs. Left + Right bias powers. Dashed lines show fit models, representing shot noise that scales in rate with detector volume with a shot noise quantum of $\varepsilon = 0.68 \pm 0.38$ meV, comparable to the aluminum superconducting bandgap.}
\end{figure}

Repeating this measurement of the correlated phonon noise amplitude, we are able to plot the correlated noise measured in the detector as a function of time (see Fig. \ref{fig:noisevtime}). As with the phonon energy resolution, we find that the correlated phonon noise drops with time, causing the resolution improvement with time.

We additionally find that the TES bias power (i.e. the power applied to the TES to keep it in transition) increases substantially over time (see Fig. \ref{fig:noisevtime}). We assume that this increase over time is essentially entirely due to a decreasing parasitic power over time, as shifts in bias power due to a cooling dilution fridge base (mixing chamber, MC) temperature should be very small (see supplementary material section \ref{appendix:mc_cooling}). As with the correlated noise, the shift in the bias power over time is approximately four times larger in the 4 mm detector than in the 1 mm detector.

As shown in Fig. \ref{fig:noisevtime} right, the correlated noise and bias power are found to be linearly related for all times in both detectors and thus share an identical functional dependence on time. This strongly suggests that both observations are caused by the same underlying mechanism.

To gain insight with a simple benchmark model, a random process in which a quantized energy deposition $\varepsilon$ occurs with an average rate $R(t)$ that scales with the detector thickness and $t^{-\kappa}$ will produce a parasitic power and excess shot noise of 
\begin{eqnarray}
\label{eqn:par_power}
    P_{1, par} = \varepsilon R(t) = \varepsilon R_0 t^{-\kappa} \\
    P_{4, par} = \varepsilon  4R(t) = \varepsilon 4 R_0 t^{-\kappa}
\end{eqnarray}
\begin{eqnarray}
\label{eqn:shot_noise}
    S_{1,ex} = 2 \varepsilon^2 R(t) = 2 \varepsilon^2 R_0 t^{-\kappa} = 2 \varepsilon P_{1,par} \\
    S_{4,ex} = 8 \varepsilon^2 R(t) = 8 \varepsilon^2 R_0 t^{-\kappa} = 2 \varepsilon P_{4,par} 
\end{eqnarray}

We plot the excess correlated noise and bias power as a function of time and find good agreement with this model (an exponential trend is not a good fit to the data). We find $\varepsilon = 0.68 \pm 0.38$ meV, and $\kappa = 0.635 \pm 0.009$. While the characteristic energy scale $\varepsilon$ seems to remain constant over time, there is no reason to assume that the shot noise events all have an identical energy. As discussed in supplementary material section \ref{appendix:shot_spectra}, $\epsilon$ is readily generalizable to processes which have a distribution of energy depositions, where $\varepsilon = \frac{<E^2>}{<E>}$. So long as the distribution is not e.g. double peaked, $\varepsilon$ can still be thought of as a characteristic energy scale of the underlying phonon burst process.

For steep exponential and power law spectra, $\varepsilon$ is on the order of the low energy cutoff to the spectrum. This offers a potential explanation for why the measured $\varepsilon$ is comparable to the aluminum superconducting bandgap energy, 2 $\Delta_{Al} \approx 360 \mu$eV, below which athermal phonons do not break Cooper pairs in the aluminum phonon collection fins of our sensors.

\begin{figure}
\includegraphics[width=1\columnwidth]{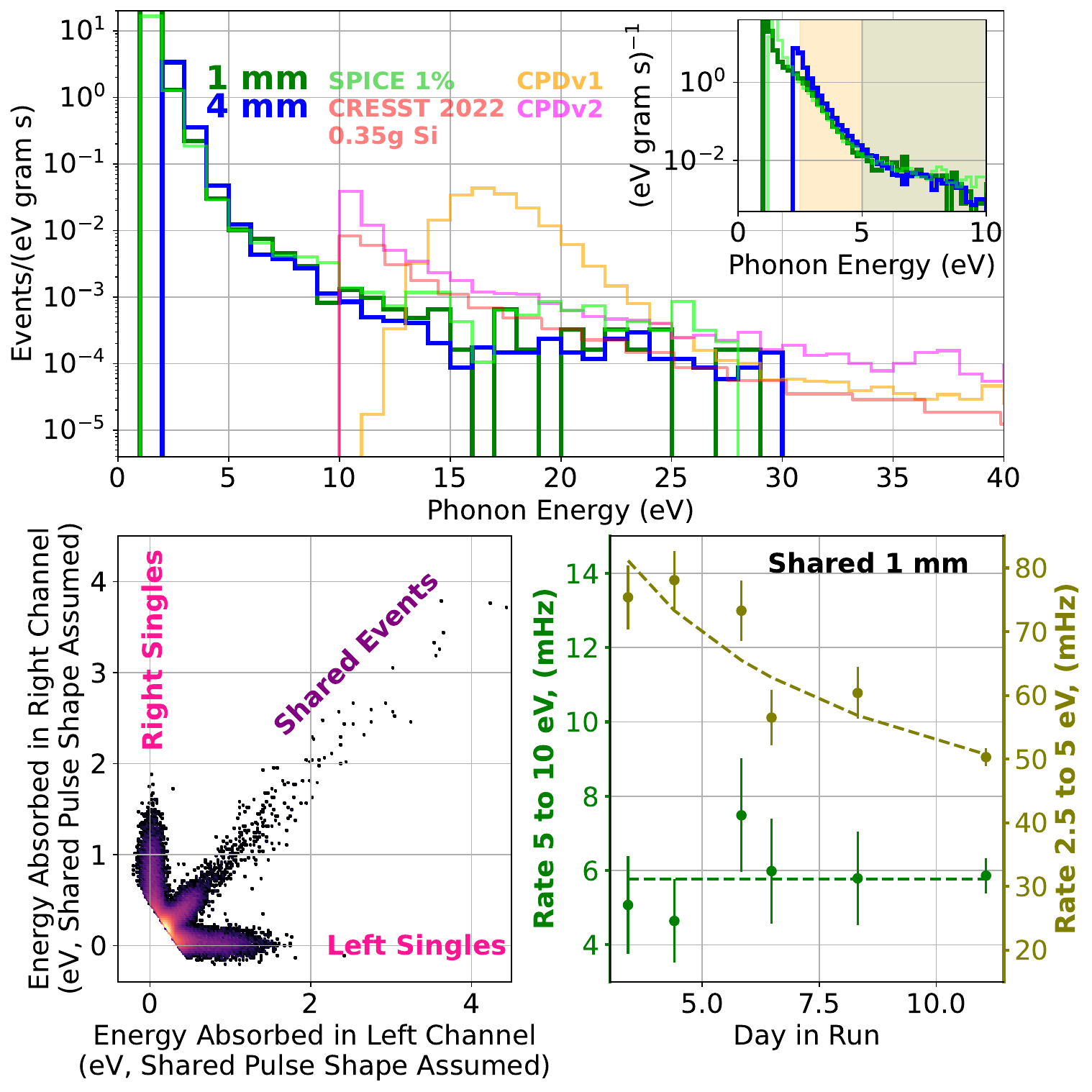}
\caption{\label{fig:backgrounds} (Top) The mass normalized shared background rates in the 1 mm and 4 mm detectors, with backgrounds measured in the SPICE 1\%\cite{TwoChannelPaper, TwoChannelLimits}, CPDv1\cite{finkPerformanceLargeArea2021, CPDLimits}, CPDv2 (previously unpublished), and CRESST 0.35g Si \cite{angloherLatestObservationsLow2022} detectors. Broadly, all six detectors seem to observe the same backgrounds normalizing for mass. Note that as the CPD and CRESST detectors are one channel detectors which cannot reject singles\cite{TwoChannelPaper}, the increased background rates near threshold could be due to singles, as well as noise triggers or additional LEE backgrounds. For clarity, the 1 mm, 4 mm, and SPICE 1\% detectors have their spectra cut off at 30 eV to remove saturated events from cosmic rays and radioactive backgrounds which begin to appear at these energies. (Top insert) Detail of the 1 mm, 4 mm, and SPICE 1\% detector backgrounds. The orange and green bands show the energy ranges in which time dependence was measured (see bottom right). (Left Bottom) Background event energy partitioning between left and right channels, showing single and shared events. Note the energy reconstruction assumes a shared pulse shape in each channel. (Right bottom) Rate of shared events over time in the 1 mm detector in the shaded 2.5-5 eV and 5-10 eV bins in the center top figure. Dashed line shows the (weighted) average rate in the 5-10 eV range, and a power law fit in the 2.5-5 eV bin.}
\end{figure}

\section{Above Threshold Shared Backgrounds}

In addition to this phonon burst shot noise from \textit{below} threshold events, we observe a high rate of \textit{above} threshold background events (often called the ``Low Energy Excess,'' or LEE\cite{adariEXCESSWorkshopDescriptions2022, LEEReview}). To investigate these LEE events, we recorded 2 or 12 hour periods of continuous background data following each calibration run, and triggered this dataset offline using an optimum filter optimized to search for events with a channel energy partition and pulse shape consistent with that of substrate absorbed calibration photons as in Ref. \cite{TwoChannelLimits}. In Fig. \ref{fig:backgrounds} bottom left, a scatter plot shows the energy deposited in each channel without a constraint on the partition. Identical to observations in Ref. \cite{TwoChannelPaper}, the LEE backgrounds split into two classes: ``shared'' events with an approximately equal channel partition and a pulse shape identical to substrate absorbed photons, and ``single'' events where nearly all energy is deposited in a single phonon sensor channel. Singles have a pulse shape with a fast rise, but a slow fall consistent with an energy deposition that saturates a subset of the athermal phonon sensors that are read out in parallel within a channel.\cite{TwoChannelPaper} Each event is then fit with a multi-channel optimum filter whose pulse shape and channel partition are fixed to the average of each LEE class (singles left, singles right, shared), and classified according to lowest $\Delta \chi^2$.\cite{TwoChannelPaper, TwoChannelLimits}

Similar to our observations of correlated phonon noise, we see that the rate of above-threshold shared background events scales with the detector thickness, as it does for below threshold phonon shot noise (the y-axis has been normalized by substrate mass, proportional to detector thickness). However, the rate of these high energy above threshold shared events is consistent with being constant with time on these short timescales. At lower energies, where statistics are better, the rate of shared LEE events does appear to decrease over time.

As in Ref. \cite{TwoChannelPaper}, our singles appear to originate within the aluminum QET films, and we additionally observe that their rate decreases with time (see further discussion in supplementary materials section \ref{appendix:singles}). These observations give additional support to the proposal that dislocation mediated relaxation of thermal stress in these aluminum QET films is responsible for these singles. \cite{TwoChannelPaper, AlRelaxation}

\section{Discussion}

We observe that both the correlated phonon noise level and the shared LEE backgrounds scale with the detector substrate thickness. From this scaling alone, we conclusively reject the hypotheses that our aluminum or tungsten sensor films constitute the dominant source of shared phonon bursts \cite{AlRelaxation}. Likewise, the phonon bursts are unlikely to originate from the top and bottom polished 1 cm$^2$ silicon faces of the substrate. 

This thickness scaling observation is compatible with a phonon burst processes whose rate scales with the diced sidewall area, the mass, or the volume of the substrate. To discriminate between these hypotheses, we would ideally measure shared LEE and correlated noise rates in detectors with different volume to sidewall ratios, while keeping all other detector characteristics fixed. Unfortunately, this study with strict extraneous variable control has not been done. 

However, we have characterized the LEE rate in two large area Cryogenic Photon Detectors (CPDs) \cite{CPDLimits, finkPerformanceLargeArea2021}. These large area cylindrical substrates (diameter: 76.2 mm, thickness 1 mm) have $\sim \times 8$ larger volume to sidewall ratio than the 1 cm$^2$ detectors discussed to this point. We observe good agreement with volume scaling, and find that when scaling by sidewall area, both noise and above threshold backgrounds differ between these two sets of detectors by around an order of magnitude. While the differences in holding method, fabrication, handling, and storage history detract from our ability to control extraneous variables, we believe that the evidence for volume scaling is compelling (see the supplementary materials \ref{appendix:mass_vs_surface_area} for additional discussion of these systematics).

Volume scaling suggests that defects in the detector substrate are responsible for LEE and excess phonon noise observations. For example, the relaxation of neutron induced crystal damage discussed in Refs. \cite{DefectLEE, NeutronDamage} could create phonon bursts consistent with our observations. The strong rate vs. time since cooldown dependence we observe in our noise data, for example, could be explained by assuming that meV-scale gaps between different energy levels in defects are thermally populated at 300 K, and then relax to lower energy states at mK temperatures. The observation of time dependence in above threshold LEE rates, both here and previously\cite{angloherLatestObservationsLow2022}, is more difficult to explain given their eV scale energies are much larger than room temperature energy scales. As discussed in Ref. \cite{DefectLEE}, an avalanche-like process where the relaxation of a meV-scale state triggers the release of many more such states is one possible mechanism by which the relaxation of thermally populated states can trigger eV scale events.

Since the shared LEE phonon bursts we observe have been shown to originate in the silicon substrate, it is likely that they would also occur in other devices constructed on silicon substrates such as superconducting qubits and MKID detectors. Phonon bursts produced in the silicon substrate would be expected to create quasiparticles in the device's superconductors, contributing to the excess equilibrium quasiparticle density long observed in these devices. To estimate the importance of this quasiparticle generation mechanism compared to other sources, we assume that the reduced quasiparticle density $x_{qp} = n_{qp}/n_{cp} = n_{qp}/(4\times10^6 \mu \mathrm{m}^{-3})$\cite{MartinisSavingQubits} in an aluminum superconductor follows the dynamical equation
\begin{equation}
    \frac{d x_{qp}}{dt} = - r x_{qp}^2 + g
\label{eq:qpdyn1}    
\end{equation}
where $r \approx (20 \mathrm{ns})^{-1}$ is the quasiparticle recombination rate and $g$ is the normalized quasiparticle generation rate \cite{MartinisSavingQubits, RadiationQubit, QPDynamicsQubit}. We assume that the quasiparticle dynamics in the superconductor are dominated by recombination as in Ref. \cite{RadiationQubit}.

Modeling the device as a silicon substrate with an aluminum superconducting region (including qubits, KIDs and ground planes) on the surface of the chip, we assume for simplicity that LEE events are evenly distributed through the bulk silicon chip and collected uniformly in the aluminum. This produces an average normalized quasiparticle generation rate of
\begin{equation}
<g>= \frac{<\rho_p> }{2 \Delta_{Al} n_{cp}} \frac{V_{Si}}{V_{Al}}
\end{equation}
where $<\rho_{p}>$ is the average LEE power density that we estimate as O(1fW/(1 cm$^{2}\times$ 1 mm) from Fig. \ref{fig:noisevtime} bottom left, $V_{Si}$ and $V_{Al}$ are the volumes of the silicon substrate and aluminum superconductor respectively. Assuming the temporal variation in $g$ is small compared to its average value $<g>$, Eq. \ref{eq:qpdyn1} can be Taylor expanded to zeroth order to find the reduced equilibrium density
\begin{equation}
    <x_{qp}> = \sqrt{\frac{<g>}{r}}= \sqrt{\frac{<\rho_p> }{2 \Delta_{Al} n_{cp}} \frac{V_{Si}}{V_{Al}} \frac{1}{r}}
\label{eq:qpdyn2a}    
\end{equation}
For the aluminum KID on a silicon substrate described in Ref. \cite{TemplesKID}, we expect a residual density of $x_{qp} \approx 4 \times 10^{-6}$ using this estimate, which compares favorably to the $x_{qp} = 4.6\times 10^{-6}$ that they observed in their KID resonator. This technique can also be used to estimate the residual quasiparticle density caused in qubits (taking into account both the $\mathcal{O}$(100 nm) thick qubit and ground plane, which cover nearly all the surface of the chip), and gives $x_{qp} \approx 10^{-8} - 10^{-7}$, roughly in line with observations for modern non-gap-engineered qubits\cite{HotQPs}. Our observations also agree with the phonon origin \cite{CorrelatedQubit} and reduction in rate over time \cite{QuasiparticleFreeSeconds, CorrelatedQubit} previously observed for such bursts in superconducting quantum devices constructed on silicon substrates.

%TC:ignore
\section{Acknowledgments}

This work was supported in part by DOE Grants DE-SC0019319, DE-SC0025523 and DOE Quantum Information Science Enabled Discovery (QuantISED) for High Energy Physics (KA2401032). This material is based upon work supported by the National Science Foundation Graduate Research Fellowship under Grant No. DGE 1106400. This material is based upon work supported by the Department of Energy National Nuclear Security Administration through the Nuclear Science and Security Consortium under Award Number(s) DE-NA0003180 and/or DE-NA0000979. Work at Lawrence Berkeley National Laboratory was supported by the U.S. DOE, Office of High Energy Physics, under Contract No. DEAC02-05CH11231. Work at Argonne is supported by the U.S. DOE, Office of High Energy Physics, under Contract No. DE-AC02-06CH11357. W.G. and Y.Q. acknowledge the support by the National High Magnetic Field Laboratory at Florida State University, which is supported by the National Science Foundation Cooperative Agreement No. DMR-2128556 and the state of Florida.

The authors have no conflicts to disclose. The data that support the findings of this study are available from the corresponding author upon reasonable request.

%\nocite{*}
\bibliography{aipsamp}% Produces the bibliography via BibTeX.

\appendix

\section{Phonon Collection Efficiency}
\label{appendix:pce}

\begin{figure}
\includegraphics[width=1\columnwidth]{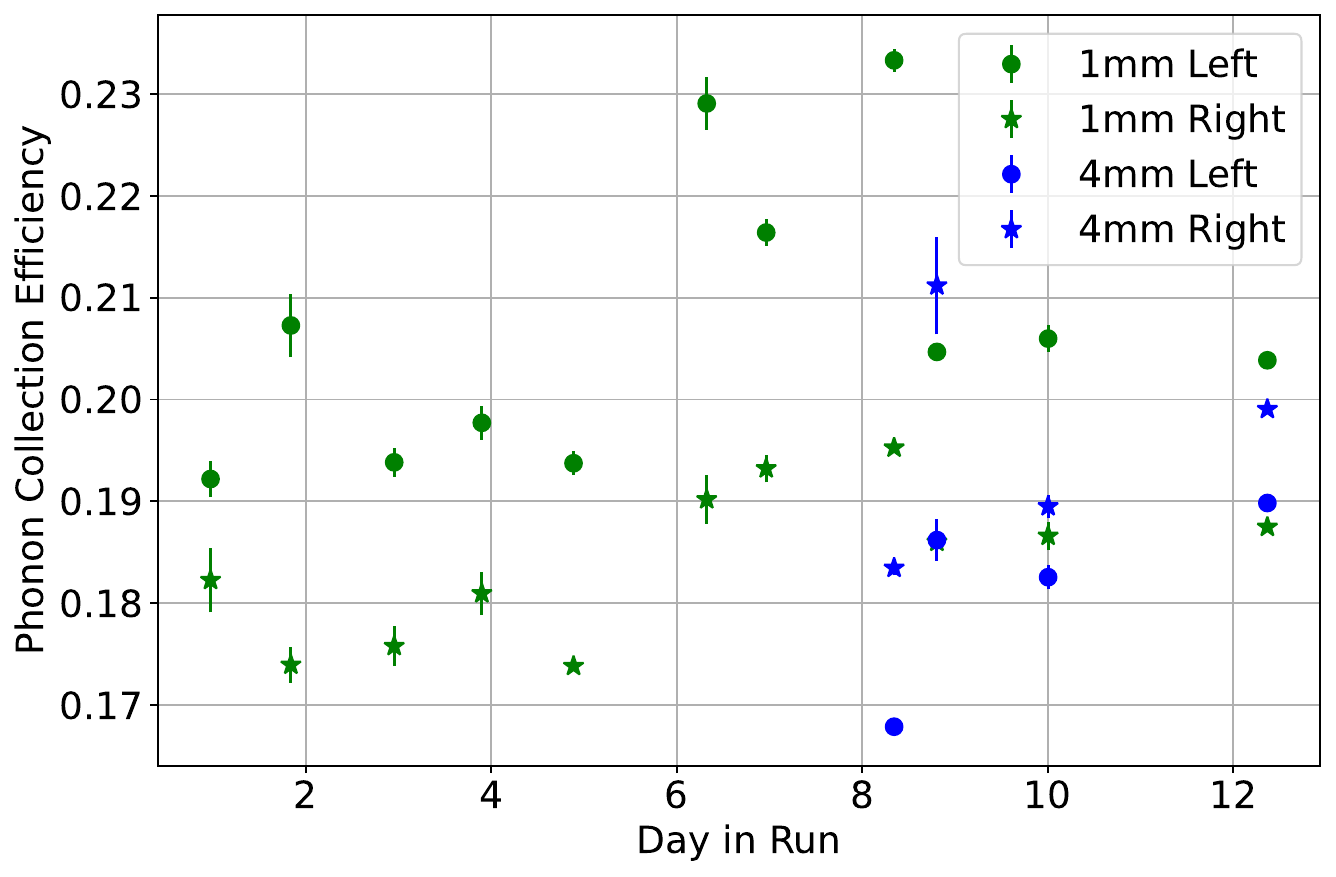}
\caption{\label{fig:pce} Phonon collection efficiencies for fit templates, measured from the photon calibrations, as a function of time. These efficiencies are roughly constant over time, meaning that the resolution improvements over time are due to improvements in the detector noise environment. 1 $\sigma$ error bars are shown.} 
\end{figure}

To monitor the detectors' phonon collection efficiency over time, we measure the responsivity $\partial P/ \partial I$ of each sensor using IV sweeps and $\partial I/\partial V$ measurements at the operating point \cite{irwinTransitionEdgeSensors2005}. We then calculate an average current-domain trace for events where a single photon is absorbed in the detector's phonon system $i(t)$. Combining this with the responsivity, we can calculate the phonon collection efficiency
\begin{eqnarray}
    \epsilon = \frac{E_{TES}}{E_\gamma} = \frac{1}{E_\gamma} \int p(t) dt = \frac{1}{E_\gamma} \int \mathcal{F}^{-1} \Big( \frac{\partial P}{\partial I} \mathcal{F}(i(t)) \Big) dt
\end{eqnarray}
where $E_\gamma$ is the energy of the photon hitting the detector and $\mathcal{F}$ and $\mathcal{F}^{-1}$ are Fourier and inverse Fourier transforms, respectively. We find that our phonon detectors' phonon collection efficiencies remain roughly constant over time, albeit with significant scatter (see Fig. \ref{fig:pce}). This variation is larger than the statistical errors on our phonon collection efficiency.

We interpret this variation as originating in fluctuations in our detectors' $\partial P/ \partial I$ over time due to time varying environmental vibrations. Our $\partial I/\partial V$ and $IV$ measurements attempt to characterize the detector in the most representative quiescent state, i.e. without any vibrational loading, and do so by being very selective. We are more accepting in our calibration events in an effort to get better statistics, allowing events with somewhat higher vibrational loading to be considered for analysis. This results in the $\partial P/\partial I$ derived from our $\partial I/\partial V$ and $IV$ measurements being slightly different from the ``true'' $\partial P/\partial I$ for events measured during the calibration, causing the reconstructed energy absorbed for the average calibration pulse to be skewed and a phonon collection efficiency measurement to be slightly systematically biased. As the level of vibration loading and therefore the size of this effect changes dataset to dataset, this would cause our measured phonon collection efficiency to vary somewhat between different datasets.

\section{Effects of Mixing Chamber Cooling}
\label{appendix:mc_cooling}

As discussed in \cite{irwinTransitionEdgeSensors2005}, a simple TES at a temperature $T$ can be thought of as a heat capacity, $C$, which is heated by bias power, $P_{bias}(t)$, parasitic power sources, $P_{par}(t)$, and the signal, $P_{signal}$. It is cooled via thermal coupling to the bath, and thus the non-linear differential equation that defines its thermal dynamics is
\begin{equation}
    C(T) \frac{dT}{dt} = P_{bias}-P_{cool}(T, T_{bath})+P_{par}(t)+P_{signal}(t)
\end{equation}
In the absence of signal excitation and with slowly varying $P_{par}$, the steady state temperature is
\begin{equation}
     P_{bias}= P_{cool}-P_{par}
\end{equation}.
The exact functional form of $P_{cool}$ depends upon the nature of the thermal impedance(s) between the detector and the thermal bath. 

Our detector housing is connected to our MC (where the bath temperature is measured) with 3 high force gold plated copper to gold plated copper joints, and our detector is thermally connected to our housing with 2 gold wire bonds. A back of the envelope estimate of the total thermal impedance of this chain \cite{Ekin2006} suggests that this thermal impedance is very small compared to the electron-phonon decoupling within the TES itself. We therefore expect that the phonon temperature of the TES is approximately the measured mixing chamber bath temperature.

To confirm that this expectation is accurate, we can measure how $P_{bias}$ varies with $T_{bath}$. If the thermal impedance is dominated by electron-phonon coupling in the TES $K_{ep}$, we expect
\begin{equation}
    P_{cool}= K_{ep}(T^{5}-T_{bath}^{5})
\end{equation}
and thus 
\begin{equation}
     P_{bias}= [K_{ep}T^{5} + P_{par}]- K_{ep}T_{bath}^{5}
\end{equation}. 
Measurements in our setup match this expectation. By contrast, poor electronic coupling across an electronic joint should lead to bias power shifts scaling as $T_{bath}^{2}$.

Over the 12 days of this run, the MC temperature and phonon bath temperature drop from 20 mK to 6 mK, both of which are significantly below the 50 mK transition temperature, $T_{c}$, of the TES. Due to the $n=5$ power law scaling, this bath temperature shift is responsible for $< 1\%$ shift in bias power, much less than the measured bias power shift.

\section{Data Quality and Single/Shared Discrimination Cuts}
\label{appendix:cuts}

As in Ref. \cite{TwoChannelPaper}, we apply quality cuts to all data series to ensure the detector is operating stably at a well characterized bias point. Additionally, we apply two special purpose cuts: a cut to ensure that traces with a large event in them are not used to construct noise estimates, and a cut which discriminated between shared and singles events in our background data.

\begin{figure}
\includegraphics[width=1\columnwidth]{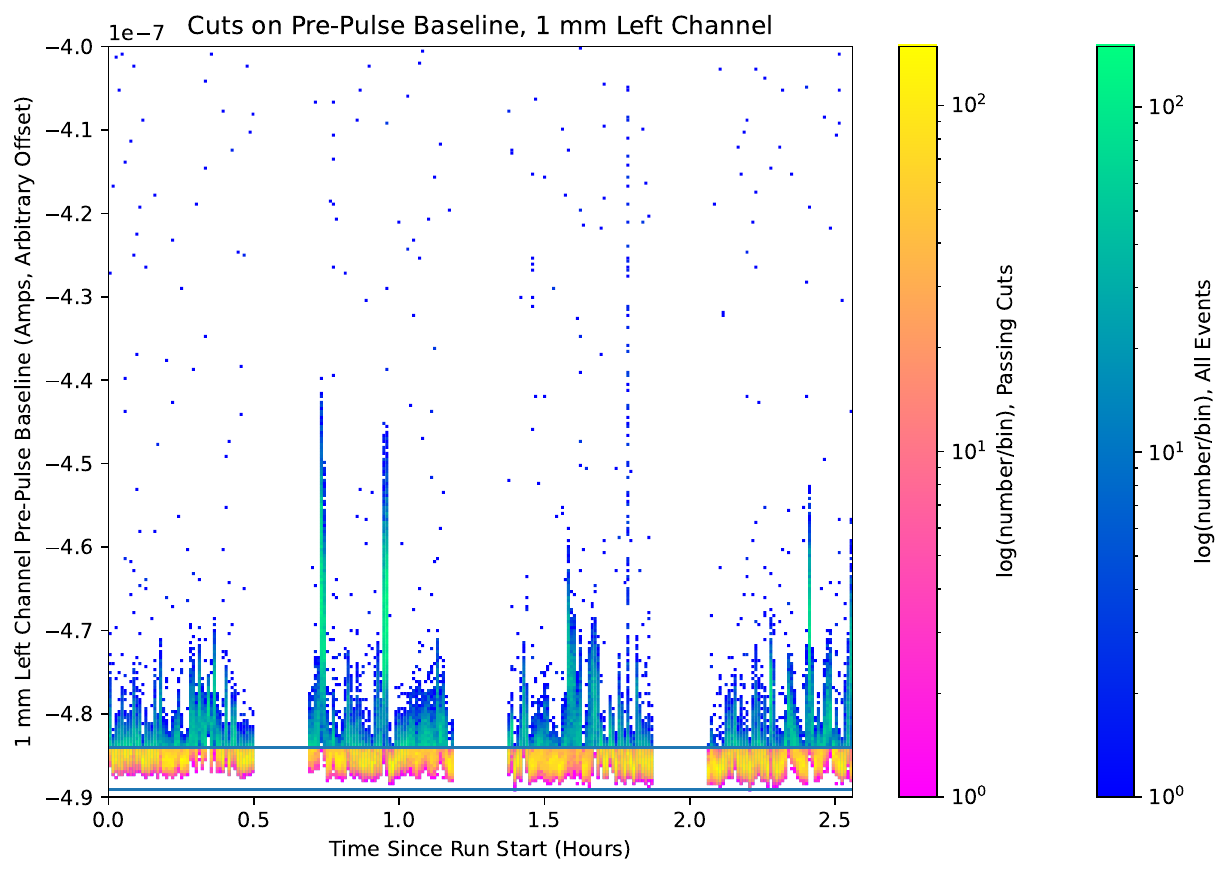}
\caption{\label{fig:baseline} The baseline cut in a representative 2 hour dataset on the 1 mm detector's left channel. The approximately minute scale variations in the detector baseline are due to shifting vibration noise in the detector, which appears at roughly intervals and intensities in all datasets. The scattered points with very high baselines are events where a pulse occurs in the pre-pulse region (e.g. pileup). In the intervals with no points, we stop taking continuous data, and characterize the detector state by taking IV and dIdV measurements to characterize the TES state and photon calibration data to characterize the detector response.} 
\end{figure}

Briefly, we cut on pre-pulse baseline, slope (pre-pulse baseline minus post-pulse baseline), and a low frequency $\chi^2$ statistic (only considering frequencies below 50 kHz), using the same general philosophy as in Ref. \cite{TwoChannelPaper}. Unlike in Ref. \cite{TwoChannelPaper}, we see significant vibration coupled noise around several hundred Hz (see Fig. \ref{fig:baseline}), and use the pre-pulse baseline and slope cuts to remove events in which this noise dominates. Our low frequency $\chi^2$ cut primarily removes pileup events, and events which were triggered at the incorrect time.

For the noise analysis, we apply an additional cut intended to remove traces with large pulses that might otherwise dominate the noise. We do this by finding the largest magnitude (positive or negative) event in the each channel's trace by using an unconstrained one channel OF, and only passing traces for which the largest event is negative in both channels, i.e. the largest event is consistent with a statistical fluctuation rather than a true event. In this sense, we select traces which are consistent with being dominated by pure noise.

\begin{figure}
\includegraphics[width=1\columnwidth]{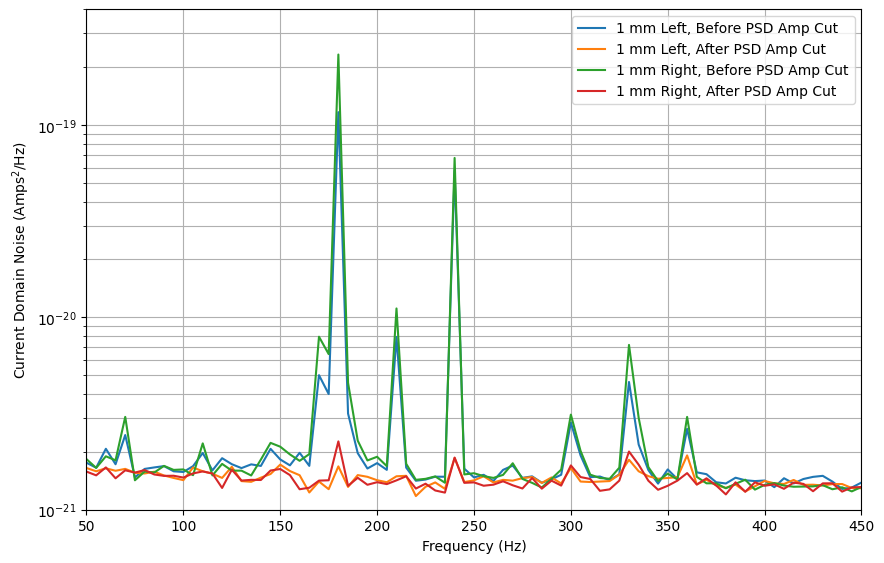}
\caption{\label{fig:psd_amp_onoff} Detail of the PSD of the 1 mm left and right channels with and without the PSD amplitude cut applied. Note that between the vibration induced peaks, the PSD amplitude cut does not change the broadband phonon shot noise (see text). Also note that the vibration-coupled noise causing the sharp peaks appears to couple more strongly to the right channel in some peaks.}
\end{figure}

\begin{figure}
\includegraphics[width=1\columnwidth]{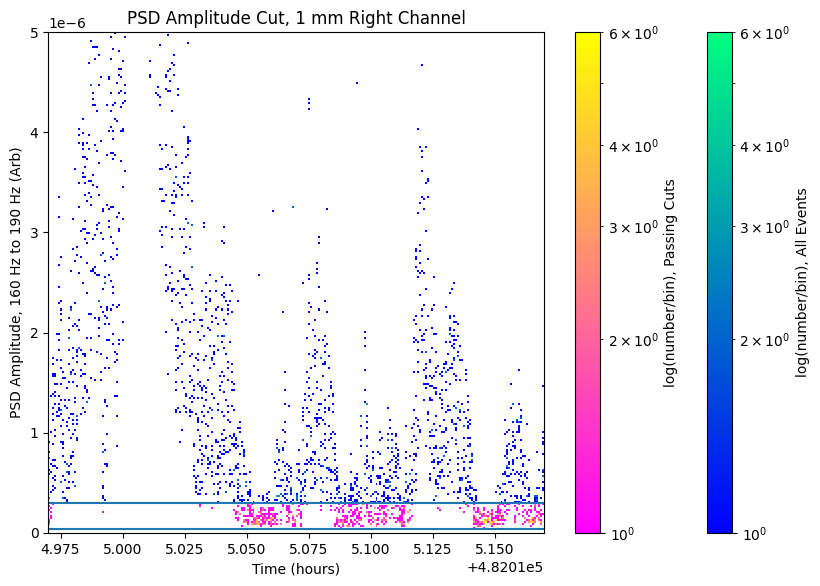}
\caption{\label{fig:psd_amp_time} Average current domain noise (arbitrary scale) in the Right channel of the 1 mm detector between 160 and 190 Hz as a function of time, showing events passing (pink) and failing (blue) the ``PSD Amplitude'' cut in this frequency range.}
\end{figure}

For the noise analysis, we also apply a more selective ``vibration amplitude'' cut to reduce time varying vibration-induced peaks in the observed PSD and CSD which were not cut by our relatively loose baseline and slope cuts (see Fig. \ref{fig:psd_amp_onoff} and Fig. \ref{fig:psd_amp_time}). These narrow resonance vibration-induced peaks (Q $\sim 10^4$) can be directly excited by mechanically thumping the fridge, and slowly decrease in average magnitude when the pulse tube is temporarily turned off (we cannot run for more than approximately 10 minutes in this configuration, and therefore take data with the pulse tube on). We also observe that these peaks are power coupled: they disappear when the TES is operated in a superconducting or normal state so that its insensitive to power signals. Interestingly, the right channel of the 1 mm detector appears to be significantly more strongly coupled to environmental vibrations than the left channel. We interpret this to mean that the mechanism injecting power into our TESs does not always couple through the detector's phonon system (which would equally couple to the two channels), and instead at least partially couples directly to individual channels. One possible explanation would be that the electrical wire bonds connecting to each channel vibrate and deposit energy directly into the TESs, with different damping constants for the wire bonds to each channel.

To perform a more selective cut of these high-vibration periods of time, we average the Fourier transform of individual traces in a relatively narrow ($\sim$30 Hz wide) frequency range around the primary vibration peak and cut events for which the average in this bin exceeds a threshold. By varying the level of this threshold, we can more or less aggressively remove periods of high vibration noise. Because of the relatively low passage of this cut, we only apply it during the noise analysis. The findings in this paper do not depend on whether or not this vibration amplitude cut is applied, e.g. the measured broad band phonon noise is the same with or without the vibration amplitude cut, as is the shift in the broadband phonon noise over time.

For the background analysis, we discriminate singles from shared events by comparing the $\chi^2$ values for 2x1 OFs with different event typologies (as in Ref. \cite{TwoChannelPaper}). For example, we select singles in the right channel by requiring that a right singles 2x1 OF has a lower $\chi^2$ value than either the best fit left singles 2x1 OF or the best fit shared 2x1 OF.

\section{Uncorrelated Noise Systematics}
\label{appendix:uncorrelated_noise}

\begin{figure}
\includegraphics[width=1\columnwidth]{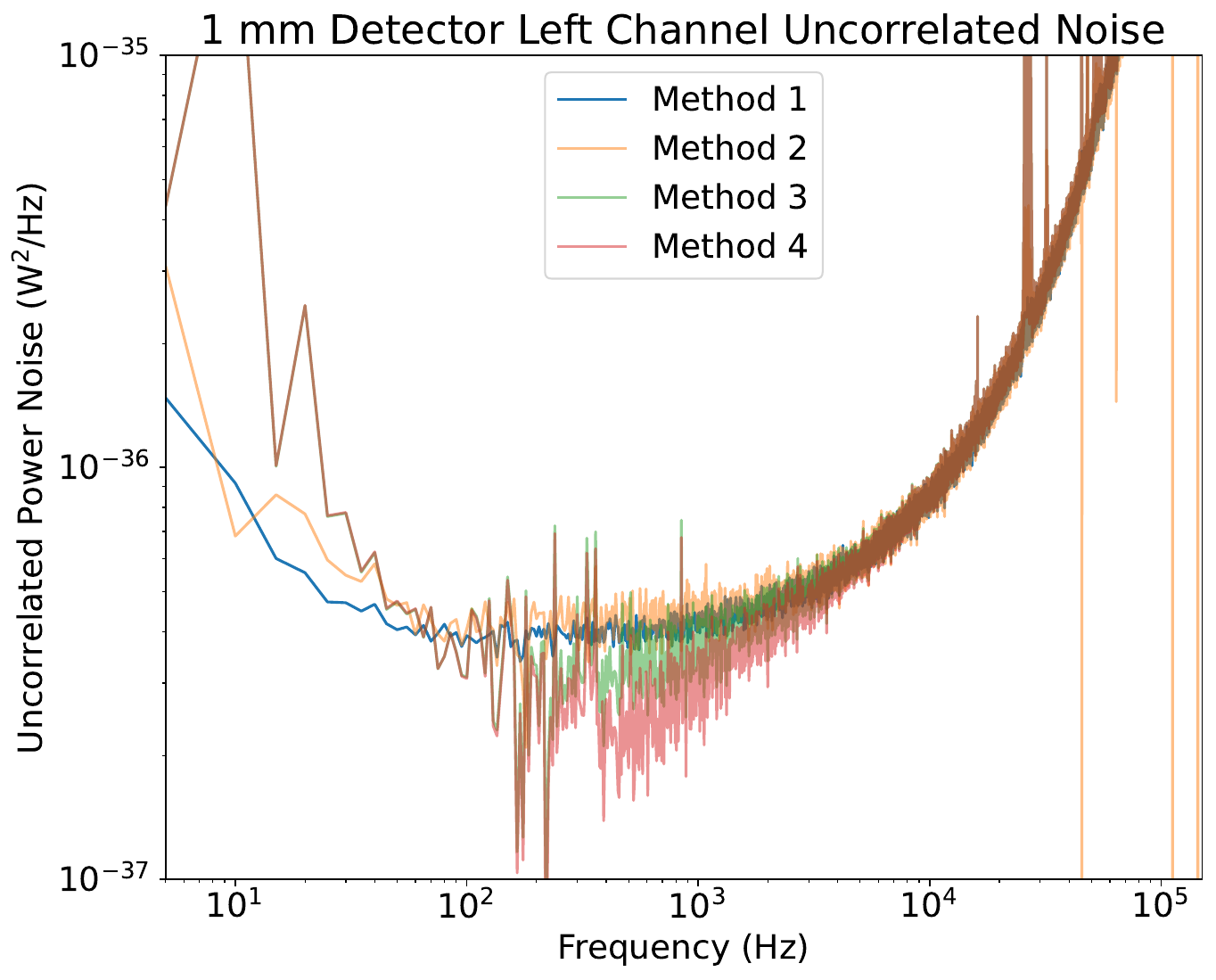}
\caption{\label{fig:uncorrelated_comparison} Uncorrelated noise in the left channel of the 1 mm detector, reconstructed using four different methods (see text). In Fig. \ref{fig:noise}, we use Method 1. Methods 3 and 4 are on top of each other at lower frequencies, producing a brown appearing line.}
\end{figure}

As in Ref. \cite{TwoChannelPaper}, we subtract the correlated noise from the total noise to find the uncorrelated noise in Fig. \ref{fig:noise}. We do this subtraction using four methods, finding consistent results (except at frequencies dominated by vibration noise where the consistency depends upon the strictness of the vibration cuts):
\begin{enumerate}
    \item We make the assumption that the uncorrelated noise in the left and right channels of the same detector is identical, and that the correlated noise couples to the two channels with a constant $k(f)$ that can vary as a function of frequency, to account for differences in channel energy partition for real events and environmental vibrations
    \begin{eqnarray}
        S_{ll}(f) = U(f) + C(f) k(f)\\
        S_{rr}(f) = U(f) + C(f)/k(f) \\
        S_{lr}(f) = C(f) \\
        j(f) = \frac{S_{ll} - S_{rr}}{S_{lr}} = k(f) - 1/k(f) \\
        k(f) = \frac{1}{2}\Big( \sqrt{j(f)^2 + 4} - j(f) \Big) \\
        U(f) = S_{ll}(f) - S_{lr}(f) k(f)
    \end{eqnarray}
    Here, $U(f)$ and $C(f)$ are the uncorrelated and correlated noise terms, and $S$ is the CSD with on diagonal elements $S_{ll},S_{rr}$ and off diagonal elements $S_{lr}$.
    \item We assume that both channels couple identically to all correlated noise sources (i.e. the k in method 1 is constrained to be 1 at all frequencies), so
    \begin{eqnarray}
        S_{ll}(f) = U_l(f) + C(f) \\
        S_{rr}(f) = U_r(f) + C(f) \\
        S_{lr}(f) = C(f) \\
        U_l(f) = S_{ll}(f) - S_{lr}(f)
    \end{eqnarray}
    This constraint is problematic in 2 ways. First, as discussed in Sec. \ref{appendix:cuts}, we find that vibrational noise has a channel coupling coefficient that is different from that of calibration photons. Thus, without extremely strict vibration cuts this model fails at vibration frequencies. Secondly, we find that optical photons have a slight asymmetry in their channel couplings.
    
    \item We model the correlated phonon noise by fitting a three time constant (two independent fall times, one shared rise time) exponential pulse model to the photon calibration data in the power domain, and by transforming this analytic pulse model to the frequency domain to get templates $t_{ll}(f)$, $t_{lr}(f)$, $t_{rr}(f)$.
    \begin{eqnarray}
        S_{ll}(f) = U_l(f) + m t_{ll}(f) \\
        S_{rr}(f) = U_r(f) + m t_{rr}(f) \\
        S_{lr}(f) = m t_{lr}(f)
    \end{eqnarray}
    where $m$ is an arbitrary scale factor we measure by fitting $t_{lr}(f)$ to $S_{lr}(f)$.

    \item Similarly to method 3, we model the correlated phonon noise by averaging photon calibration events, and by transforming this averaged data-driven template to the power and frequency domain to get templates $t_{ll}(f)$, $t_{lr}(f)$, $t_{rr}(f)$.
    \begin{eqnarray}
        S_{ll}(f) = U_l(f) + m t_{ll}(f) \\
        S_{rr}(f) = U_r(f) + m t_{rr}(f) \\
        S_{lr}(f) = m t_{lr}(f)
    \end{eqnarray}
    where $m$ is an arbitrary scale factor we measure by fitting $t_{lr}(f)$ to $S_{lr}(f)$. Methods 3 and 4 differ in that 3 uses an analytic model of the pulse shape, whereas 4 uses a data-driven averaged calibration pulse as the correlated noise model. Methods 3 and 4 have significantly fewer degrees of freedom compared to 1 and 2 since the shape of the correlated noise spectrum is fixed by the measured photon calibration pulse shape.
    
\end{enumerate}
See Fig. \ref{fig:uncorrelated_comparison} for a graphical comparison of these methods. 

\begin{figure}
\includegraphics[width=1\columnwidth]{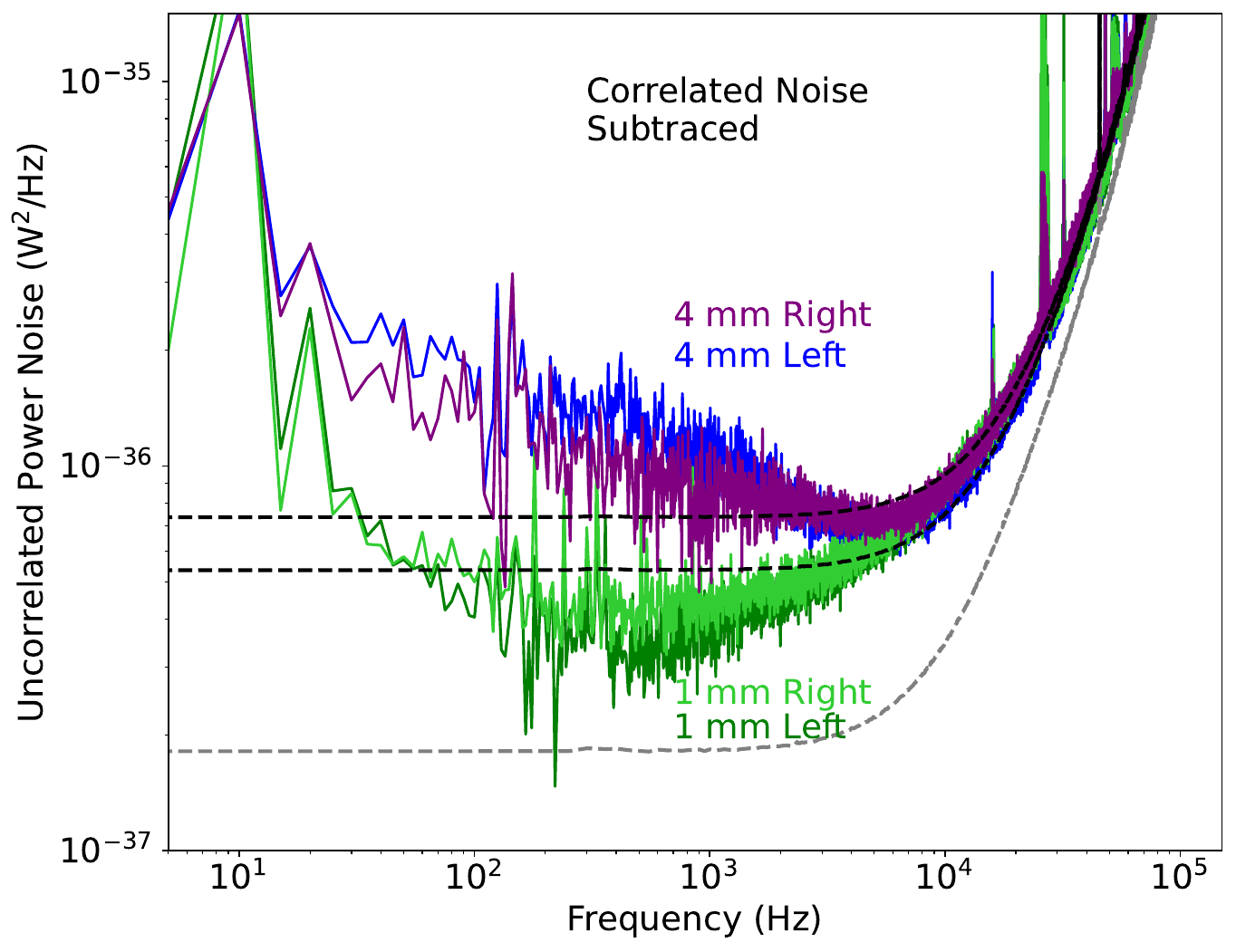}
\caption{\label{fig:uncorrelated} Uncorrelated noise in each channel, measured using method 4 (see above). Grey dashed line shows modeled TES noise for the 1 mm left channel, all other channels have similar modeled noise spectra. Black dashed lines show TES noise + an arbitrary broadband constant, showing that no channel seems compatible with this simple model proposed in Ref. \cite{TwoChannelPaper}. }
\end{figure}

The uncorrelated noise obviously deviates from the TES noise + flat shot noise model we proposed in Ref. \cite{TwoChannelPaper}, and deviates in seemingly opposite ways in the 1 mm and 4 mm detectors (Fig. \ref{fig:uncorrelated}). Several possibilities might explain this deviation from the expected form:
\begin{itemize}
    \item Variation in the partition of energy between the channels due to position dependence of subthreshold LEE will decrease the amount of correlated shot noise and increase the uncorrelated shot noise. Furthermore, this additional uncorrelated shot noise will have a phonon pulse shape roughly matching the shape of the uncorrelated spectrum seen in the 4mm device. 
    
    The measured partition distribution for the optical photon calibration is consistent with a $\sim$ 4\% fractional position dependence. However, this likely overestimates the true position dependence for uniformly distributed events throughout the bulk of the substrate since the optical calibration photons are preferentially absorbed on the phonon sensor instrumented surface. In the 4 mm device, the excess measured uncorrelated noise is consistent with a $8 \times 10^{-37}$ W$^2$/Hz flat noise possibly from localized relaxation in the aluminum fins plus additional shot noise with a phonon shape due to a 2\% fractional energy position dependence. 

    The excess shot noise seen in the 1 mm device above the phonon collection bandwidth ($\sim$ 2.9 kHz) but below the TES sensor pole ($\sim$ 10 kHz, dashed grey line in the upper right subplot of Fig. \ref{fig:noise}) can not be understood with such a simple model. Potentially, models where the position dependence changes both the pulse shape and the energy partition between the 2 channels could model the rise in excess uncorrelated noise found at higher frequencies in the 1 mm device. This explanation is not entirely self-consistent; we both expect and see more position dependence in the 1 mm thick device in the photon calibration, yet see less uncorrelated noise at all frequencies.

    \item Our power-to-current responsivity model ($dP/dI$) might be inaccurate. We use standard models for this responsivity \cite{ThreePoledPdI, irwinTransitionEdgeSensors2005} which accurately predict current-domain pulse shapes for energy impulses in our TESs from photons \cite{TESVeto}, giving us some reassurance that we are accurately modeling TES responsivity. However, our QETs (which have a different structure from TES photon sensors) could have additional thermal dynamics that do not couple as expected to our $\partial I / \partial V$ excitations. For example, ``third'' ($\sim$ 800 Hz) poles seen in our $\partial I / \partial V$ could couple to current through the TES in addition to the thermal couplings proposed in Ref. \cite{ThreePoledPdI}. Directly probing our TESs' $dP/dI$ through e.g. exciting our TESs with a low energy ($\sim 300$ meV) photon source could probe this potential systematic.

    \item We do not attempt to model the Internal Thermal Fluctuation Noise (ITFN) \cite{ThreePoledPdI} caused by stochastic energy fluctuations between the independent thermal degrees of freedom in our three pole TES dynamical model.  Given that our observed noise (even after ``decorrelation'') dominated our modeled TFN noise by close to an order of magnitude, we find it unlikely that this ITFN noise due to additional heat capacities could be dominant. 
\end{itemize}

Ultimately, we cannot pinpoint the origin of this effect, and are therefore reluctant to draw strong conclusions about the origin of our uncorrelated noise. We do note that the uncorrelated noise is roughly equal in magnitude for both 1 mm and 4 mm channels around 3-5 kHz, where neither correlated phonon noise nor TES noise should be dominant. This suggests that a common noise source (e.g. originating within the sensors themselves) causes this noise in all channels, however, without an accurate noise model, we feel that this observation is inconclusive.

\section{Shot Noise From Energy Spectra}
\label{appendix:shot_spectra}

Generically, we can extend the results of Eqs. \ref{eqn:par_power} and \ref{eqn:shot_noise} to account for arbitrary event spectra $\frac{d R}{d E}(E)$ by integrating
\begin{eqnarray}
    P_{par} = \int_{\varepsilon_l}^\infty E \frac{d R}{d E}(E) dE \\
    S_p = 2\int_{\varepsilon_l}^\infty E^2 \frac{d R}{d E}(E) dE
\end{eqnarray}
where $\varepsilon_l$ is the lowest energy event the detector is sensitive to which is set by the superconducting bandgap, $2 \Delta_{Al}$). Combining these two qualities yields
\begin{eqnarray}
    \epsilon = \frac{S_p}{2 P_{par}} = \frac{<E^2>}{<E>} = \frac{\int_{\varepsilon_l}^\infty E^2 \frac{d R}{d E}(E) dE}{\int_{\varepsilon_l}^\infty E \frac{d R}{d E}(E) dE}
\end{eqnarray}

For an exponential event distribution
\begin{eqnarray}
    \frac{d R}{d E}(E) = \frac{R_0}{\varepsilon_0} e^{-E/\varepsilon_0} 
\end{eqnarray}
we find
\begin{eqnarray}
    P_{par} = R_0 e^{-\varepsilon_l/ \varepsilon_0} (\varepsilon_l + \varepsilon_0) \\
    S_p = 2 R_0 e^{-\varepsilon_l/ \varepsilon_0} (2\varepsilon_0^2 + 2 \varepsilon_0 \varepsilon_l + \varepsilon_l^2) \\
    \frac{S_p}{2 P_{par}} = \frac{2 \varepsilon_0^2 + 2 \varepsilon_0 \varepsilon_l + \varepsilon_l^2}{(\varepsilon_l + \varepsilon_0)}
\end{eqnarray}
Fitting a line to the $S_p$ vs. $P_{par}$ data, we would fit $2 \varepsilon_0$ or $\varepsilon_l$ for $\varepsilon_l >> \varepsilon_0$ and $\varepsilon_l << \varepsilon_0$ respectively.

For a power law distribution with a pivot energy $E_1$,
\begin{eqnarray}
    \frac{d R}{d E}(E) = R_1 \bigg( \frac{E}{E_1} \bigg)^{- \eta}
\end{eqnarray},
we find
\begin{eqnarray}
    P_{par} = \frac{R_1 \varepsilon_l^2}{2 - \eta} \bigg(\frac{\epsilon_l}{E_1}\bigg)^{2-\eta} \\
    S_p = 2 \frac{R_1 \varepsilon_l^{3}}{3 - \eta} \bigg(\frac{\epsilon_l }{E_1}\bigg)^{3-\eta} \\
    \frac{S_p}{2 P_{par}} = \epsilon_l \frac{2 - \eta}{3 - \eta} 
\end{eqnarray}
for $\eta > 3$.

Note that while here we integrate up to infinity, these integrals are essentially unaffected by integrating up to a finite cutoff $\varepsilon_h$ as long as $\varepsilon_h >> \varepsilon_l$ and $\varepsilon_h >> \varepsilon$ for exponential distributions and $\eta > 3$ for the power law model. (For $\eta \leq 3$ in the power law model, the integrals will not converge.) Equivalently, we can say that the ultimate shot noise we measure does not strongly depend on how we cut high energy events, as long as the underlying distribution of events are steep and we cut very high energy events which do not belong to the underlying exponential or power law distribution (e.g. saturated events from cosmic rays). For scale, our $\varepsilon_l$ (and $\varepsilon$ in the exponential distribution scenario) should be around the aluminum bandgap $2\Delta_{Al} \approx $ 0.4 meV, and our high energy cutoffs will be on the order of $\sigma_p$, i.e. hundreds of meV, decisively putting us in the regime where the severity of our high energy event cut does not strongly set the correlated noise level we measure. We have explicitly tested this by accepting events for the noise analysis with a varying energy cutoff, and have confirmed that the ultimate correlated noise level does not vary with different energy level cuts, as long as the very largest events are cut.

\section{Singles Energy Spectrum, Saturation Effects, and Rate vs. Time}
\label{appendix:singles}

As in Ref. \cite{TwoChannelPaper}, we have studied the pulse shape of the singles pulses observed in the background dataset and compared them to the aluminum direct hit events in the calibration dataset, and conclude as we did in Ref. \cite{TwoChannelPaper} that the background singles events originate within the aluminum films.

In addition to these pulse shape studies, we have measured the rate of singles events over time. To do this, we trigger singles events on a single channel optimal filter, using the averaged pulse shape of above threshold singles events as a template. To estimate the energy deposited in the TES, the best fit current amplitude is multiplied by the measured $\partial P/\partial I (f)$ as in Refs. \cite{TwoChannelPaper, TESVeto}. As described in the main text, we discriminate these events from shared events by comparing the $\chi^2$ values for two channel, one amplitude (2x1) optimal filters assuming both singles and shared responses.

\begin{figure}
\includegraphics[width=1\columnwidth]{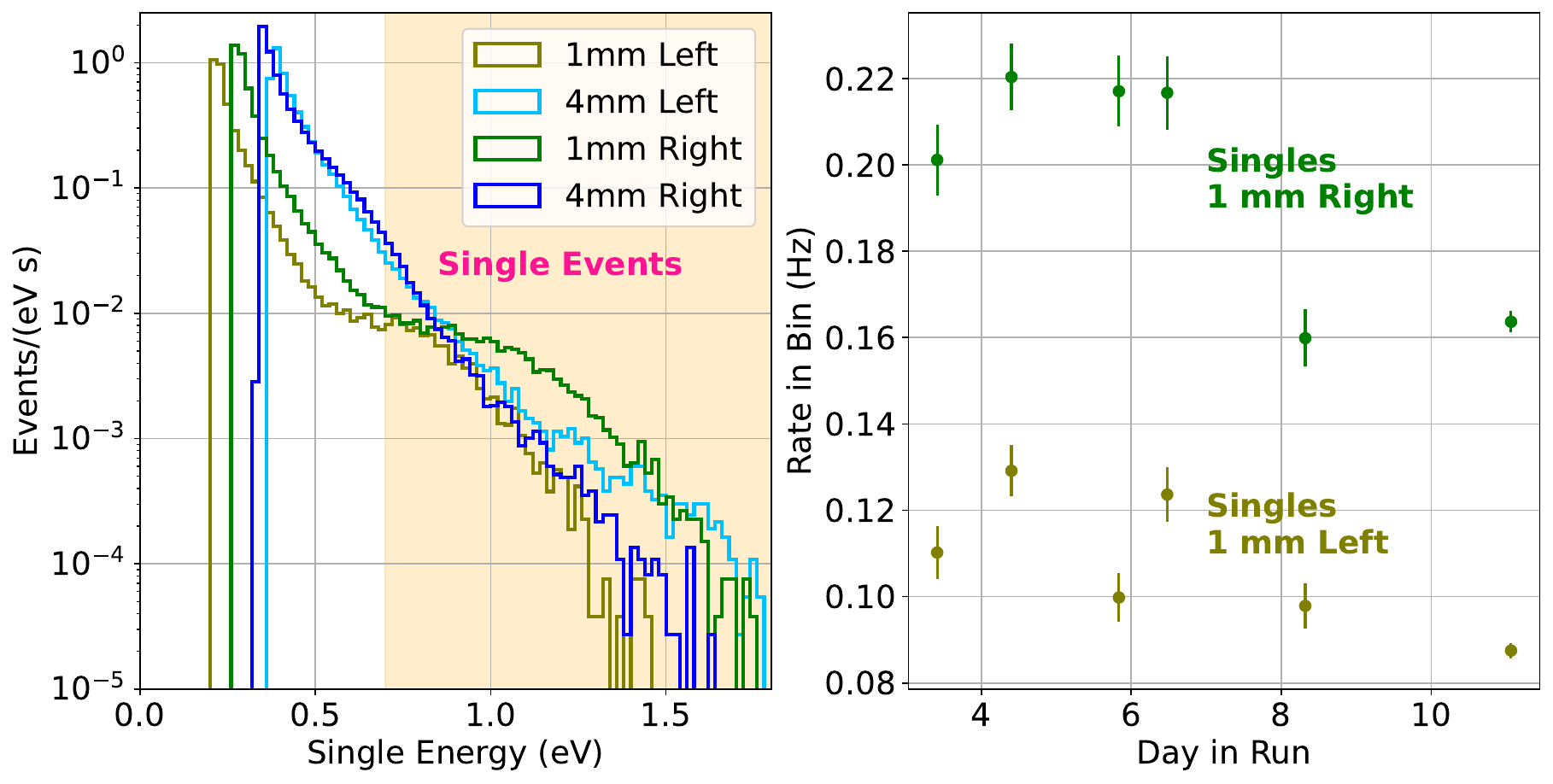}
\caption{\label{fig:singles} (Left) Spectra of singles events observed in the 1 mm and 4 mm detectors. Energies correspond to an absorbed energy in the TES. (Right) Singles rate between 0.7 and 1.8 eV over time, corresponding to the orange band in the left panel. } 
\end{figure}

Comparing the left and right singles spectra for the 1 and 4 mm devices, we see differences in the rate and energy scale of singles observed in different channels. Presumably, some of these differences are due to saturation effects, which would be expected to distort spectra at different energies and relative to each other. As we discuss in Ref. \cite{TwoChannelPaper}, our singles appear to deposit energy at a single point within a specific QET's aluminum fins, with energy unevenly shared between the several TESs closest to this event. These different TESs receive different amounts of energy, and therefore saturate at different energies, leading to an artificially broadened and inherently energy dependent pulse shape. We observe evidence of energy dependence in these pulse shapes, and by fitting a pulse model composed of a delta-function and saturated component find that the average singles pulse shape becomes less saturated and more delta-like at lower energies, consistent with this model.

However, some channel-to-channel variation might remain even if the saturation related effects could be compensated for. In principle, some film-to-film variation of singles energy and rate might be expected, given these singles appear to originate with the aluminum QET films. This might be due to, for instance, the different thermal histories of the 1 mm and 4 mm aluminum films, as thicker substrates would be expected to remain cooler during aluminum deposition.

\section{Comparison and Analysis of Event Rates and Noise over Time}
\label{appendix:global_over_time}

\begin{figure}
\includegraphics[width=1\columnwidth]{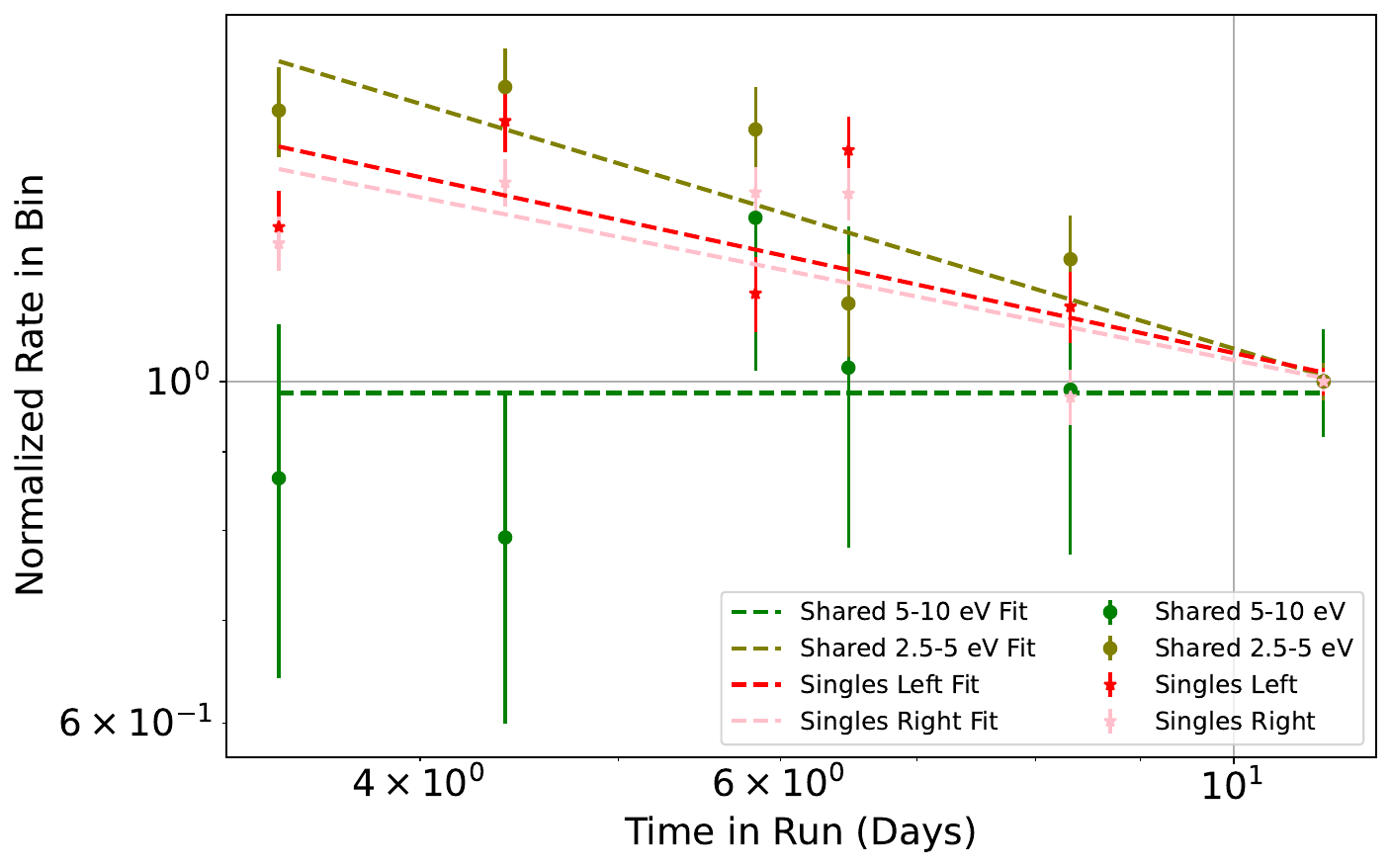}
\caption{\label{fig:shared_singles_vs_time} Shared and single rates vs. time in the 1 mm detector, normalized to the rate in the last dataset. Power law fits are shown with dashed lines. For shared events between 5 and 10 eV, we show a constant with time model.} 
\end{figure}

\begin{figure}
\includegraphics[width=1\columnwidth]{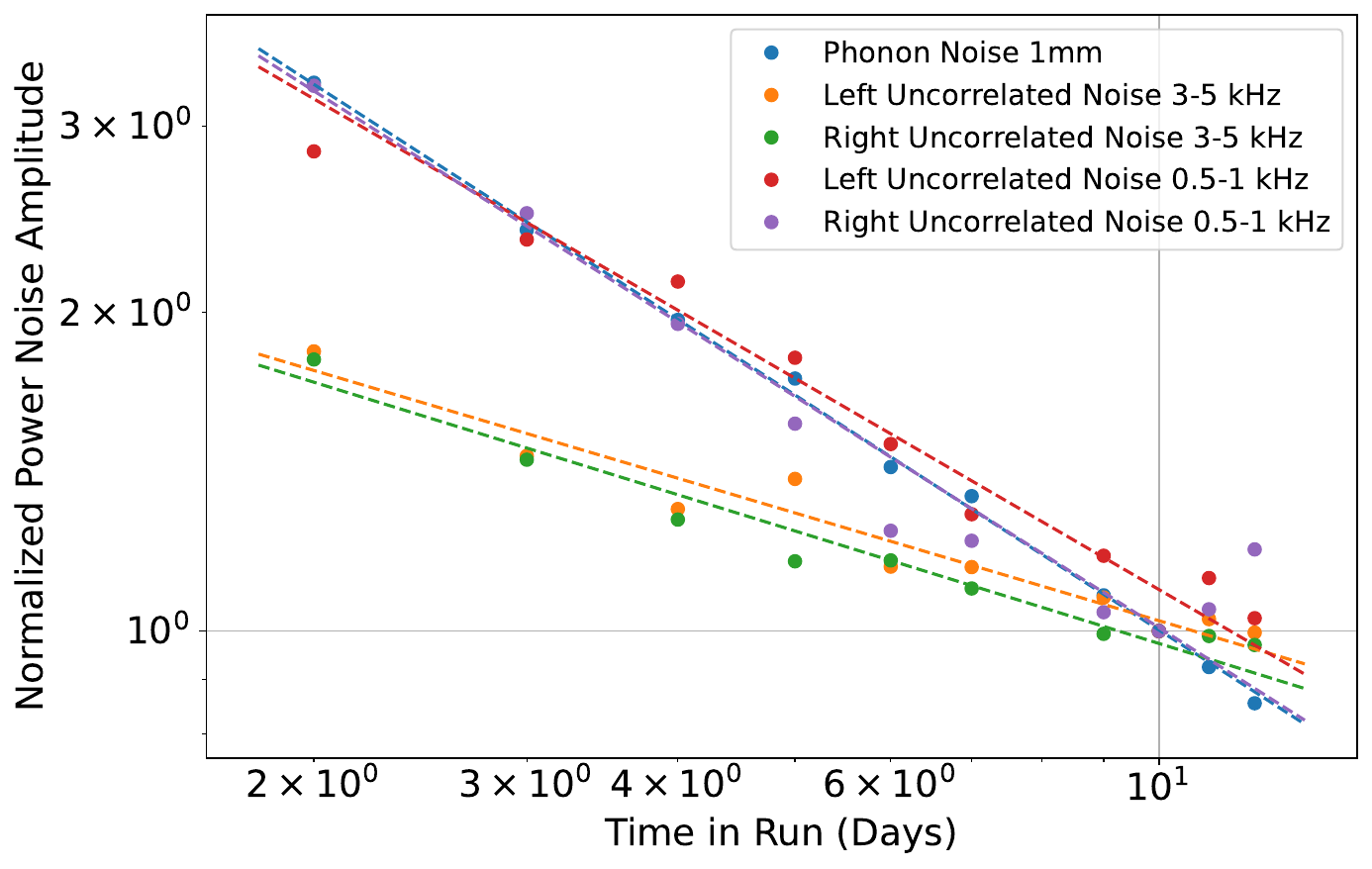}
\caption{\label{fig:cor_uncor_vs_time} Normalized correlated and uncorrelated noise vs. time models in the 1 mm device, with power law fits shown. The color of the dashed fit lines corresponds to the color of the data points shown in the legend.} 
\end{figure}

\begin{figure}
\includegraphics[width=1\columnwidth]{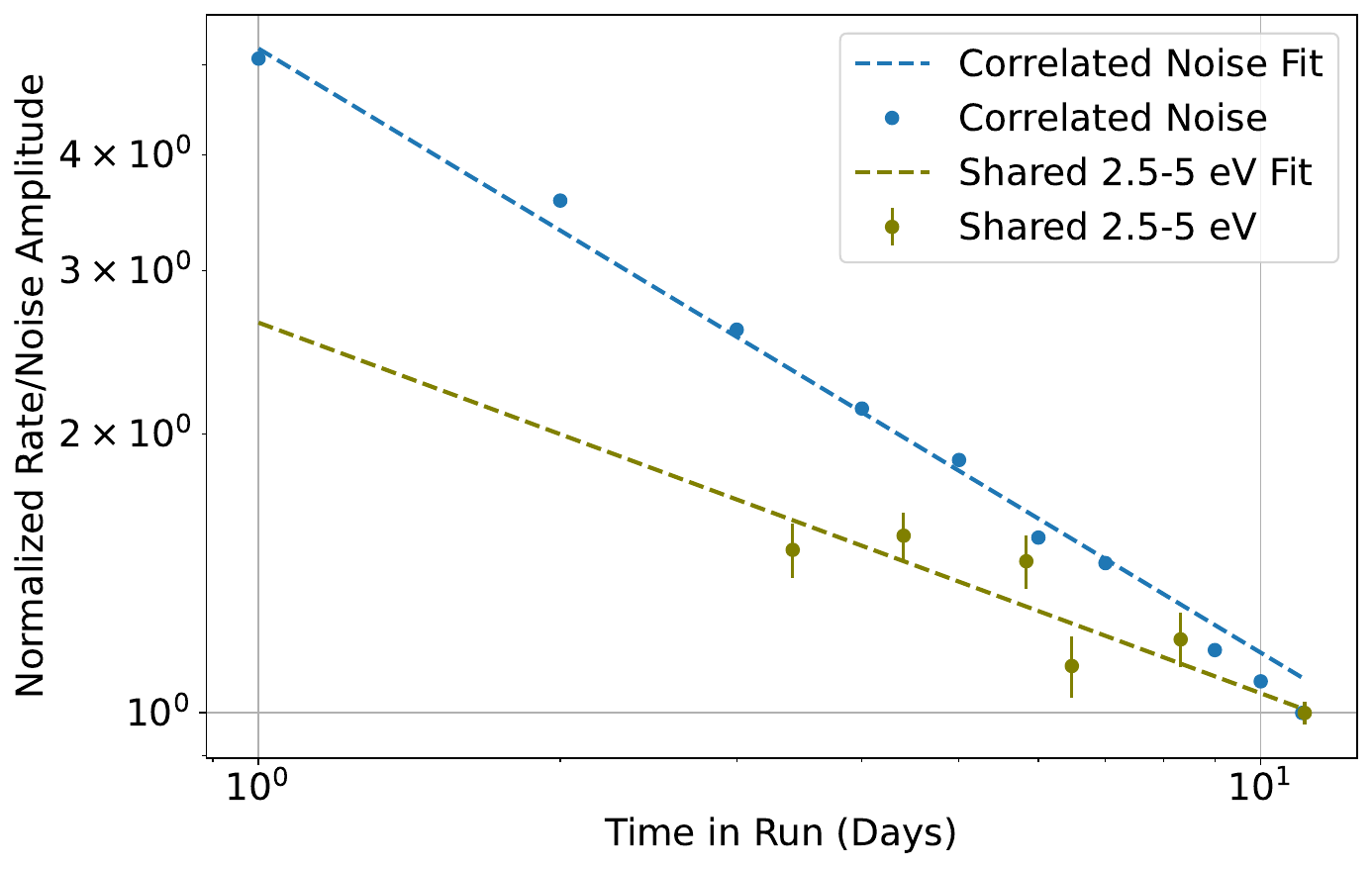}
\caption{\label{fig:shared_cor_vs_time} Normalized correlated noise and shared backgrounds in the 1 mm device between 2.5 and 54 eV, with power law fits shown. These power law fits are statistically inconsistent.} 
\end{figure}

In this letter and supplementary material, we measure the noise and background rate as a function of time in several channels in the 1 mm detector:
\begin{itemize}
    \item The correlated noise amplitude.
    \item The ``uncorrelated'' noise rate between 500 Hz and 1 kHz
    \item The ``uncorrelated'' noise rate between 3 and 5 kHz
    \item The shared background rate between 2.5 and 5 eV
    \item The shared background rate between 5 and 10 eV
    \item The singles background rate
\end{itemize}
To test whether these time varying backgrounds and noise terms may be related, we show fits to the data using simple power law models (see Figs. \ref{fig:shared_singles_vs_time}, \ref{fig:cor_uncor_vs_time},  \ref{fig:shared_cor_vs_time}).

\begin{table}[ht]
\begin{center}
\begin{tabular}{|c|c|} 
    \hline
    Channel Name & Power Law Constant $\kappa$\\
    \hline
    Correlated Noise & $0.635 \pm 0.008$ \\
    \hline
    Left Uncorrelated Noise 0.5-1 kHz & $0.66 \pm 0.07$ \\
    \hline
    Right Uncorrelated Noise 0.5-1 kHz & $0.73 \pm 0.05$ \\
    \hline
    Left Uncorrelated Noise 3-5 kHz & $0.34 \pm 0.03$ \\
    \hline
    Right Uncorrelated Noise 3-5 kHz & $0.35 \pm 0.02$ \\
    \hline
    Shared Backgrounds, 5-10 eV & $-0.10 \pm 0.12$ \\
    \hline
    Shared Backgrounds, 2.5-5 eV & $0.40 \pm 0.07$ \\
    \hline
    Left Singles Backgrounds & $0.29 \pm 0.09$ \\
    \hline
    Right Singles Backgrounds & $0.27 \pm 0.08$ \\
    \hline
\end{tabular}
\label{tab:fit_kappa}
\caption{Power law constants, from power law models fit to time series data in different noise and background channels in the 1 mm detector. }
\end{center}
\end{table}

Table \ref{tab:fit_kappa} shows the fit power law constant $\kappa$ for the various noise and background channels. The uncorrelated noise and singles backgrounds have statistically consistent power law constants in the left and right channels, consistent with the idea that the two channels observe the same underlying processes. 
Additionally, we see that below the primary phonon pole ($\sim$ 2.9 kHz in the 1 mm detector, dashed green line in the upper right subplot of Fig. \ref{fig:noise}) the $\kappa$ for the uncorrelated noise is statistically consistent with that of the correlated noise. This is consistent with the idea, discussed above, that position dependence in the phonon noise spectrum leads to phonon coupled ``uncorrelated'' noise. By contrast, the time dependence of the uncorrelated noise in the 3-5kHz bin has similar time dependence to that of the above threshold singles, suggesting that in this frequency range, the uncorrelated noise is dominated by shot noise from subthreshold singles LEE events.

In principle, our observed correlated noise could be due to a phonon spectrum that continues to be observable above threshold, however, comparing $\kappa$s for the correlated noise and 2.5-5 eV shared backgrounds shows statistically inconsistent time dependencies. As discussed in the main text, different mechanisms seem required to give time dependence on the meV and eV energy scales.

Finally, we note that the 2.5-5 eV and singles backgrounds appear to have statistically consistent time dependencies. However, our observation of the shared background rate scaling with the detector thickness and the pulse shape evidence for singles originating within the aluminum QET fins strongly suggest that these two event classes have different origins, and that the statistical consistency between their time dependencies is purely coincidental.

\section{Background Events and Correlated Noise Scaling With Mass vs. Surface Area, Comparison to Other Experiments}
\label{appendix:mass_vs_surface_area}

Using the 1 mm and 4 mm detectors characterized in this work, we show that both the rate of above threshold background events (i.e. LEE) and the correlated noise amplitude scale with the thickness of the detector substrate. This observation is compatible with three possible origins of such noise and backgrounds. First, these events and this noise could originate from the sidewalls of the detector, caused by e.g. saw damage from dicing. Alternatively, they could originate from the bulk volume of the detector substrate itself. Finally, the LEE rate could scale with mass of the substrate, but not originate in the volume. To our knowledge the only hypothesis that fits this third category is creep/plastic deformation of the structural wirebond foot that scales weight of the substrate. 

The first two hypotheses could be distinguished by detectors with different sidewall area to volume ratios, testing whether the noise and backgrounds scale linearly with either detector volume or sidewall area. Unfortunately, no detectors with different sidewall area to volume ratios were manufactured at the same time as the detectors used in this work, preventing us from performing a direct test with the expectation of well controlled systematic variables. We aim to fabricate such detectors and directly test sidewall area vs. detector volume scaling of noise and background event rates in future work. 

\begin{figure}
\includegraphics[width=1\columnwidth]{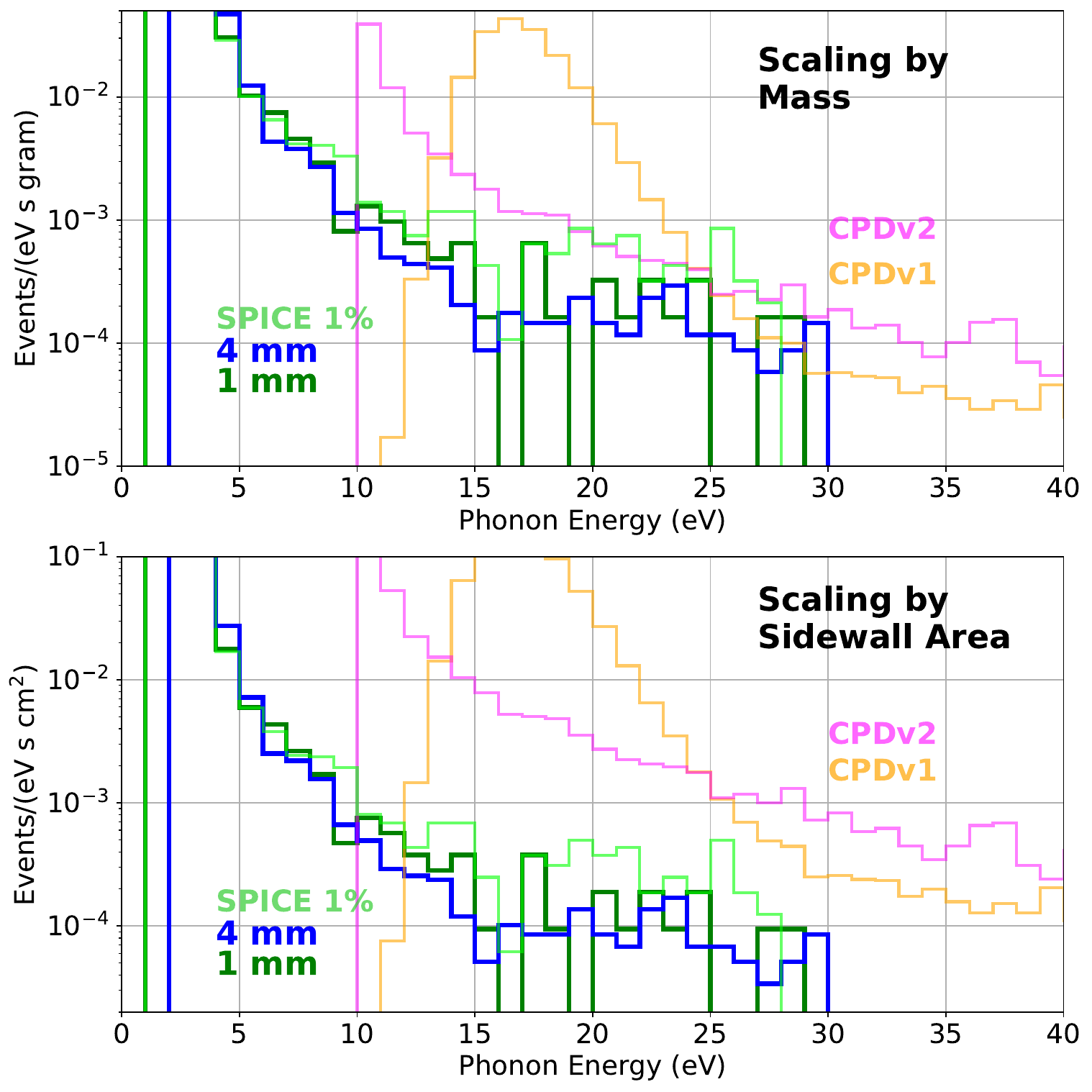}
\caption{\label{fig:mass_a_comp} The rate of events observed in the 1 mm and 4 mm detectors compared to the SPICE 1\% detector\cite{TwoChannelPaper, TwoChannelLimits}, the CPDv1 detector\cite{CPDLimits, finkPerformanceLargeArea2021}, and the CPDv2 detector (previously unpublished) using the same coloring and energy range as in Fig. \ref{fig:backgrounds}, scaling both by the mass of the detector (top) and by the sidewall area (bottom). As we do not know how the CRESST 0.35g Si detector\cite{angloherLatestObservationsLow2022} was diced (i.e. which faces are sawed sidewalls vs. polished top and bottom faces), we do not include that detector for comparison. Note that the CPDv1 and v2 detectors are single channel detectors that cannot discriminate ``singles,'' possibly contributing to the rise in events just above threshold of those detectors. At higher energies, the LEE rate appears to better match between the two detector form factors (1 cm$^2$ detectors: 1 mm, 4 mm, SPICE 1\% detectors, vs. 3 inch diameter CPD detectors) when scaling by mass as opposed to sidewall area, suggesting that the LEE causing these high energy events originates within the bulk of the detector. A number of possible systematics effects might obscure an underlying scaling, see text for further discussion.} 
\end{figure}

However, previously run detectors (the ``CPDv1'' detector in \cite{finkPerformanceLargeArea2021} and an updated, previously unpublished version of the same detector, ``CPDv2'') with different sidewall area to volume ratios \textit{are} available for comparison. We compare these two form factors of detectors (1 cm$^2$ for the 1 mm, 4 mm, and SPICE 1\% detectors vs. the 76.2mm diameter by 1mm thick CPD detectors) in Fig. \ref{fig:mass_a_comp}, and see that volume (as opposed to sidewall surface area) scaling appears to better match the data at higher energies. As the CPD detectors are single channel, higher threshold detectors, we expect that singles and noise triggers would contribute a significant number of events just above threshold, possibly explaining the difference in observed rate between the cm$^2$ detectors and CPD detectors near threshold.

Unfortunately, several potential systematic effects make direct comparison with the 1 mm and 4 mm detectors worse than a better controlled test.
\begin{itemize}
    \item \textbf{Time since cooldown:} while in this work, we do not seem to observe a change in the observed LEE rate as a function of time above 5 eV, other detectors (see e.g. Ref. \cite{angloherLatestObservationsLow2022}) have observed a clear rate vs. time dependence at 10s of eV. Because the data for CPDv1 was taken significantly later in the run than the data for the CPDv2 and cm$^2$ detectors (34 days vs. 11-12 days into the run, see table \ref{tab:dimensions_trig_comp}), rate vs. time dependencies could complicate the comparison of CPDv1 to the other detectors.
    \item \textbf{Detector holding:} The 1 mm and 4 mm detectors as well as the SPICE 1\% detector are suspended by wire bonds as described in Ref. \cite{AnthonyPetersen2024}, while the CPDv1 and v2 detectors are clamped with Cirlex and sapphire clamps respectively. As detector holding methods which induce high stress in the detector are known to create backgrounds in the detector \cite{astromFractureProcessesObserved2006, AnthonyPetersen2024}, the difference in holding methods between the two form factors of detectors could possibly add an additional LEE source in one or both sets of detectors. However, recent research by the CRESST collaboration led them to conclude that the LEE rate did not strongly vary between detectors with different clamping schemes\cite{CRESSTClampingTest} where the clamping was sufficiently weak such that it did not observavbly crack the detector substrate (as it did in Ref. \cite{astromFractureProcessesObserved2006}). This result is compatible with our observations assuming that the LEE rate scales with detector volume. 
    \item \textbf{Time since manufacture:} The time elapsed between detector fabrication and an LEE measurement could effect the measured LEE rate, for example if heating the detector substrate during e.g. photoresist baking changed the LEE rate. The estimated time between fabrication and data taking for each detector is summarized in Table \ref{tab:dimensions_trig_comp}. We see no clear relationship between the observed LEE rate and the time since fabrication, but leave open the possibility that such a systematic effect could obscure underlying trends in the data.
    \item \textbf{Single channel vs. two channel detectors:} The CPD detectors are single channel detectors, and consequently cannot discriminate singles and shared events. The spectra for the CPD detectors therefore include singles, which presumably contribute significantly to the spectrum close to threshold, where the CPD detector spectra diverge from the cm$^2$ detector spectra.
    \item \textbf{Differences in threshold and triggering:} The CPD detectors have significantly higher thresholds than the cm$^2$ detectors, presumably due in part to their phonon noise scaling with detector volume (as we observe in this work).
\end{itemize}

\begin{table*}[ht]
\begin{center}
\begin{tabular}{|c|c|c|c|c|c|c|c|} 
    \hline
    Detector Name & Dimensions & Holding & Mass (g) & Surface Area (cm$^2$) & Sidewall Area (cm$^2$) & Day in Run & Days Since Fab\\
    \hline
    This work, 4 mm & 1 cm$^2$ $\times$ 4 mm & Hanging\cite{AnthonyPetersen2024} & 0.932 & 3.6 & 1.6 & 12 & $\sim$ 90 \\ 
    \hline
    This work, 1 mm  & 1 cm$^2$ $\times$ 1 mm & Hanging\cite{AnthonyPetersen2024} & 0.233 & 2.4 & 0.4 & 12 & $\sim$ 90 \\ 
    \hline
    SPICE 1\% \cite{TwoChannelPaper} & 1 cm$^2$ $\times$ 1 mm & Hanging\cite{AnthonyPetersen2024} & 0.233 & 2.4 & 0.4 & 11 & $\sim$ 191 \\ 
    \hline
    CPD v1 \cite{finkPerformanceLargeArea2021, CPDLimits} & 7.6 cm \O $\times$ 1 mm & Cirlex clamps & 10.6 & 47.75 & 2.39 & 34 & $\sim$ 685 \\ 
    \hline
    CPD v2\footnote{Not previously published, an updated version of the detector described in Ref. \cite{finkPerformanceLargeArea2021}.} & 7.6 cm \O $\times$ 1 mm & Sapphire clamps & 10.6 & 47.75 & 2.39 & 12 & $\sim$ 355 \\ 
    \hline
    CRESST 0.35g Si 2022\cite{angloherLatestObservationsLow2022} & (2 cm)$^2$ $\times$ 0.4 mm & Copper clamps & 0.35 & 8.32 & 0.32 & $\sim$250-350 & Unknown\\ 
    \hline
\end{tabular}
\caption{\label{tab:dimensions_trig_comp} Dimensions, weight, and holding of low threshold detectors considered in this work. ``Day of Run'' indicates how long after cooldown the spectra in Figs. \ref{fig:backgrounds}, \ref{fig:mass_a_comp} were taken. ``Days since Fab'' estimates the elapsed time between the end of device fabrication and the start of datataking.}
\end{center}
\end{table*}

While we believe that it is relatively improbable that any of these systematics should dominate our backgrounds, and similarly improbable that such systematics would happen to coincidentally make backgrounds in the cm$^2$ and CPD detectors appear to be compatible with volume scaling, we cannot eliminate the possibility that our observations are due to poorly controlled systematic effects. We aim to perform a more carefully controlled test of mass vs. sidewall area scaling with a future set of detectors.

The fact that the shared above threshold LEE seems consistent with mass/volume scaling for different structural support techniques (wirebond hanging, Cirlex clamping, sapphire clamping, copper clamping) disfavors the hypothesis that the above threshold shared LEE or phonon noise are due to wirebond foot creep.

\begin{table*}[h!]
\begin{center}
\begin{tabular}{|c|c|c|c|c|c|c|} 
    \hline
    Detector Name & Correlated Noise & Day of Run & Noise / Mass & Scaled Noise / Mass & Noise / Sidewall Area & Scaled Noise / Sidewall Area\\
     & (W$^2$/Hz) & & (W$^2$/(Hz g)) & (W$^2$/(Hz g)) & (W$^2$/(Hz cm$^2$)) & (W$^2$/(Hz cm$^2$))\\
    \hline
    This work, 4 mm & $3.6 \times 10^{-36}$ & 12 & $3.8\times 10^{-36}$ & $3.8\times 10^{-36}$ & $2.3\times 10^{-36}$ & $2.3\times 10^{-36}$ \\ 
    \hline
    This work, 1 mm & $7.6 \times 10^{-37}$  & 12 & $3.2\times 10^{-36}$ & $3.2\times 10^{-36}$ & $1.9\times 10^{-36}$ & $1.9\times 10^{-36}$ \\ 
    \hline
    SPICE 1\% \footnote{This detector had degraded phonon collection efficiency and therefore lower observed noise due to accidental etching of its tungsten TESs. We correct for this by scaling the noise up by the ratio of the phonon collection efficiencies measured for this work and for the SPICE 1\% detector, i.e. (39.1\%/28.9\%) = 1.33.} \cite{TwoChannelPaper} & $1.0 \times 10^{-36}$ & 7 & $4.4\times 10^{-36}$  & $3.1\times 10^{-36}$ & $2.5\times 10^{-36}$ & $1.8\times 10^{-36}$ \\ 
    \hline
    CPD v1\footnote{Single channel detector, noise reduced by a factor of two for comparison to two channel detectors.} \cite{finkPerformanceLargeArea2021, CPDLimits} & $4.2 \times 10^{-35}$ & 24 &  $4.0\times 10^{-36}$ & $ 6.2\times 10^{-36}$ & $ 1.7\times 10^{-35}$ & $2.6\times 10^{-35}$ \\ 
    \hline
    CPD v2\footnote{Not previously published, an updated version of the detector described in Ref. \cite{finkPerformanceLargeArea2021}. Single channel detector, noise reduced by a factor of two for comparison to two channel detectors.} & 4.7 $\times 10^{-35}$ & 6  & $4.5\times 10^{-36}$ & $2.9\times 10^{-36}$ & $ 2.0\times 10^{-35}$ & $1.3\times 10^{-35}$ \\ 
    \hline
\end{tabular}
\caption{\label{tab:noise_comp} Correlated noise amplitude comparison in different low threshold phonon detectors. The two single channel CPD detectors have their noise reduced by a factor of two to correct for the noise being absorbed in a single channel, as opposed to one of two channels. The final column scales the noise to the expected noise on day 12, assuming the noise falls off with the same power law as measured here ($\kappa = 0.63$).}
\end{center}
\end{table*}

%TC:endignore

\end{document}